\documentclass{sbc2019}%

\usepackage{graphicx}
\usepackage[utf8]{inputenc}
\usepackage[misc,geometry]{ifsym} 
\usepackage{fontspec}
\usepackage{fontawesome}
\usepackage{academicons}
\usepackage{color}
\usepackage{hyperref} 
\usepackage{aas_macros}
\usepackage[bottom]{footmisc}
\usepackage{supertabular}
\usepackage{afterpage}
\usepackage{url}
\usepackage{pifont}
\usepackage{multicol}
\usepackage{multirow}
\usepackage{caption}
\usepackage{subcaption}
\usepackage{listings}
\usepackage[none]{hyphenat}
\usepackage{svg}
\usepackage{makecell}
\usepackage{trimclip}
\usepackage{placeins}

\definecolor{codegreen}{rgb}{0,0.6,0}
\definecolor{codegray}{rgb}{0.5,0.5,0.5}
\definecolor{codepurple}{rgb}{0.58,0,0.82}
\definecolor{backcolour}{rgb}{0.95,0.95,0.92}
\lstdefinestyle{mystyle}{
    backgroundcolor=\color{backcolour},   
    commentstyle=\color{codegreen},
    keywordstyle=\color{magenta},
    numberstyle=\tiny\color{codegray},
    stringstyle=\color{codepurple},
    basicstyle=\ttfamily\scriptsize,
    breakatwhitespace=false,         
    breaklines=true,                 
    captionpos=b,                    
    keepspaces=true,                 
    numbers=left,                    
    numbersep=5pt,                  
    showspaces=false,                
    showstringspaces=false,
    showtabs=false,                  
    tabsize=2
}
\setcitestyle{authoryear,open={[},close={]}}

\definecolor{orcidlogo}{rgb}{0.37,0.48,0.13}
\definecolor{unilogo}{rgb}{0.16, 0.26, 0.58}
\definecolor{maillogo}{rgb}{0.58, 0.16, 0.26}
\definecolor{darkblue}{rgb}{0.0,0.0,0.0}
\hypersetup{colorlinks,breaklinks,
            linkcolor=darkblue,urlcolor=darkblue,
            anchorcolor=darkblue,citecolor=darkblue}

\jid{JBCS}
\jtitle{Journal of Internet Services and Applications, 2024, XX:X, }
\doi{10.5753/jisa.2024.XXXXXX}
\copyrightstatement{This work is licensed under a Creative Commons Attribution 4.0 International License}
\jyear{2024}

\title[Virtualized 5G Tesbed using OpenAirInterface: Tutorial and Benchmarking Tests]{Virtualized 5G Tesbed using OpenAirInterface: Tutorial and Benchmarking Tests}

\author[Dória et al. 2024]{
\affil{\textbf{Matheus Dória}~\href{https://orcid.org/0000-0002-2278-3953}{\textcolor{orcidlogo}{\aiOrcid}}[\textbf{Universidade Federal do Rio Grande do Norte~}|\href{mailto:matheus.fagundes.067@ufrn.edu.br}{~\textbf{\textit{matheus.fagundes.067@ufrn.edu.br}}}~]}

\affil{\textbf{Vicente Sousa}~\href{https://orcid.org/0000-0003-2859-6136}{\textcolor{orcidlogo}{\aiOrcid}}~\textcolor{blue}{\faEnvelopeO}~[~\textbf{Universidade Federal do Rio Grande do Norte~}|\href{mailto:vicente.sousa@ufrn.br}{~\textbf{\textit{vicente.sousa@ufrn.br}}}~]}

\affil{\textbf{Antonio Campos}~\href{https://orcid.org/0000-0003-4350-6389}{\textcolor{orcidlogo}{\aiOrcid}}~~[~\textbf{Universidade Federal do Rio Grande do Norte}~|\href{mailto:antonio.campos@ufrn.br}{~\textbf{\textit{antonio.campos@ufrn.br}}}~]}

\affil{\textbf{Nelson Oliveira}~\href{https://orcid.org/0009-0003-9014-3725}{\textcolor{orcidlogo}{\aiOrcid}}~~[~\textbf{Universidade Federal do Rio Grande do Norte}~|\href{mailto:nelson@imd.ufrn.br}{~\textbf{\textit{nelson@imd.ufrn.br}}}~]}

\affil{\textbf{Paulo Eduardo}~\href{https://orcid.org/0009-0003-4499-4254}{\textcolor{orcidlogo}{\aiOrcid}}~~[~\textbf{Universidade Federal do Rio Grande do Norte~}|\href{mailto:paulo.eduardo.093@ufrn.edu.br}{~\textbf{\textit{paulo.eduardo.093@ufrn.edu.br}}}~]}

\affil{\textbf{Paulo Filho}~\href{https://orcid.org/0000-0002-7909-4331}{\textcolor{orcidlogo}{\aiOrcid}}~~[~\textbf{Universidade Federal do Rio Grande do Norte}~|\href{mailto:paulo.filho.071@ufrn.edu.br}{~\textbf{\textit{paulo.filho.071@ufrn.edu.br}}}~]}

\affil{\textbf{Carlos Lima}~\href{https://orcid.org/0000-0001-7933-289X}{\textcolor{orcidlogo}{\aiOrcid}}~~[~\textbf{Universidade Federal do Rio Grande do Norte~}|\href{mailto:carlos.lima.106@ufrn.edu.br}{~\textbf{\textit{carlos.lima.106@ufrn.edu.br}}}~]}

\affil{\textbf{João Guilherme}~\href{https://orcid.org/0000-0002-3092-7603}{\textcolor{orcidlogo}{\aiOrcid}}~~[~\textbf{Universidade Federal do Rio Grande do Norte~}|\href{mailto:joao.guilherme.016@ufrn.edu.br}{~\textbf{\textit{joao.guilherme.016@ufrn.edu.br}}}~]}

\affil{\textbf{Daniel Luna}~\href{https://orcid.org/0000-0003-4194-924X}{\textcolor{orcidlogo}{\aiOrcid}}~~[~\textbf{Universidade Federal do Rio Grande do Norte~}|\href{mailto:daniel.luna.088@ufrn.edu.br}{~\textbf{\textit{daniel.luna.088@ufrn.edu.br}}}~]}

\affil{\textbf{Iago Rego}~\href{https://orcid.org/0000-0002-2082-0140}{\textcolor{orcidlogo}{\aiOrcid}}~~[~\textbf{Universidade Federal do Rio Grande do Norte~}|\href{mailto:iago.diogenes.072@ufrn.edu.br}{~\textbf{\textit{iago.diogenes.072@ufrn.edu.br}}}~]}

\affil{\textbf{Marcelo Fernandes}~\href{https://orcid.org/0000-0001-7536-2506}{\textcolor{orcidlogo}{\aiOrcid}}~~[~\textbf{Universidade Federal do Rio Grande do Norte~}|\href{mailto:mfernandes@dca.ufrn.br}{~\textbf{\textit{mfernandes@dca.ufrn.br}}}~]}

\affil{\textbf{Augusto Neto}~\href{https://orcid.org/0000-0002-9936-3770}{\textcolor{orcidlogo}{\aiOrcid}}~~[~\textbf{Universidade Federal do Rio Grande do Norte~}|\href{mailto:augusto@dimap.ufrn.br}{~\textbf{\textit{augusto@dimap.ufrn.br}}}~]}

}

\begin{document}

\begin{frontmatter}
\maketitle

\begin{mail}
Leading Advanced Technologies Center of Excellence (LANCE), Universidade Federal do Rio Grande do Norte (UFRN), Campus Universitário Central da UFRN, R. das Engenharias, s/n, Lagoa Nova, Natal - RN, 59078-900, Brazil. 
\end{mail}

\begin{dates}
\small{\textbf{Received:} DD Month YYYY~~~$\bullet$~~~\textbf{Accepted:} DD Month YYYY~~~$\bullet$~~~\textbf{Published:} DD Month YYYY}
\end{dates}


\begin{abstract}
\textbf{Abstract}
\noindent The development of 5G and its evolutionary path to 6G brings virtualization as close as possible to the antennas. Native 3GPP systems are now software running at servers boosted by accelerator cards to cope with computationally intense signal processing. Meanwhile, the Radio Frequency (RF) front-end is still proprietary hardware, sheltering specific PHY-layer procedures like 
passband amplification/modulation. This approach takes advantage of the software's flexibility while keeping the complex microsecond signal processing performance from modern telecommunication systems. This paper provides tutorial material on the Core Network and Radio Access Network of OpenAirInterface (OAI) 5G stack on top of Universal Software Radio Peripheral (USRP) platforms. A set of blueprints showcases OAI's ability to provide a virtualized 5G network with different transmission capabilities and the possibility to use it with commercial mobile phones. Configuration discussions and throughput benchmark analyses follow installation and deployment instructions. Our results show that 5G prototyping using OAI and USRP frontends can lead to good reproducibility and consistent throughput.

\end{abstract}



\end{frontmatter}

\section{Introduction}
\label{sec_intro}

The rapid evolution of 5G technology has highlighted the need for flexible and cost-effective network deployment and testing solutions. 
Additionally, the virtualization of telecommunication systems is a challenging evolution movement, especially regarding the opportunity to provide an open and free software platform worldwide while offering a telecommunication system designed to face adverse conditions. The stability of the virtualized mobile network extends far beyond merely avoiding crashes; it encompasses the capability to perform reliably under various scenarios and loads. From a business point of view, downtime can frustrate users, leading to dissatisfaction and, ultimately, causing churn. Thus, the stability of a telecommunication system means economic survival.

Open Air Interface (OAI)\footnote{http://openairinterface.org} offers an open-source platform that allows researchers and professionals to implement and test 5G networks using software-defined radio (SDR) technology. This paper provides a comprehensive tutorial on setting up a virtualized 5G network using OAI and USRP, focusing on practical deployment scenarios and performance benchmarking. The primary motivation for this work is to showcase how open-source tools like OAI can revolutionize research and development in telecommunications. By providing an accessible and low-cost environment for 5G network prototyping, this tutorial offers a practical solution. With detailed instructions and performance analyses, it becomes a valuable resource for both academia and industry, fostering the replication and further exploration of the studies conducted. Additionally, this paper seeks to fill a gap in the existing literature, where many works focus on theoretical aspects or performance results without providing a step-by-step guide for setting up and conducting experiments. By including detailed analyses and solutions to common issues, we aim to increase OAI's reproducibility and practical utility for a wide range of applications, from training to technological innovation initiatives.




Thus, the research contributions of this paper are as follows:

\begin{itemize}
    \item providing a set of blueprints that showcases OAI's ability to provide a virtualized 5G network with different transmission capabilities and its use with commercial mobile phones;
    \item contributing to increase the reproducibility of 5G prototyping using the OAI and USRP RF frontends;
    \item providing a throughput benchmark analysis and a discussion of 5G RAN configuration parameters to reach OAI's maximum capacity;
    \item make all the source codes for RAN configurations and use case scenarios available in an open-source repository for global community access.
\end{itemize}

The paper is structured as follows. Section~\ref{subsec_relworks} provides an analysis of key related works; Section~\ref{sec_virtual} presents the OAI, along with the system's architecture; Section~\ref{sec_install} introduces the tutorial material for the installation and deployment process, while 
Section ~\ref{sec_TestScen} introduces the installation and deployment process and describes the tests performed, including the equipment used and their configuration parameters. Then, Section ~\ref{sec_Results} presents and discusses the results from performance and benchmark tests. Lastly, Section~\ref{sec_conclusion} concludes with final remarks and future works.

\subsection{Related Works}
\label{subsec_relworks}

OpenAirInterface (OAI) is an open-source project that implements 3rd Generation Partnership Project (3GPP) technology in general-purpose computing hardware and Software Defined Radio (SDR) cards. It allows deploying and operating 4G Long-Term Evolution (LTE) or 5G New Radio (NR) networks at a low cost. 
Projects with open-source code can be adapted for different use cases and deployments, positioning them as an ideal industrial or academic research platform. The OAI Software Alliance (OSA) is a non-profit consortium that fosters a community of industrial and academic research collaborators. The OSA also developed the open source OAI public license, allowing contributors to implement their patented technology without giving up their intellectual property rights \citep{KALTENBERGER2020107284}.

In 2020, some interesting works addressed the use of OAI in virtualization, prototyping, and testbeds. In \citep{KALTENBERGER2020107284}, the functionalities, capabilities, and continuous evolution of the OAI were considered, describing the OAI's 5G implementation efforts, both on the radio network side and the core network side. The authors described how OAI fits into 5G Standalone (SA) and Non-Standalone (NSA) scenarios, providing implementation details about the OAI's 5G gNB. EPC Control and User Plane Separation (CUPS) and 5GC are discussed, considering the OAI implementation and the 3GPP vision. The authors also described preliminary results and some of the 5G NR features deployed in an external test environment at Eurecom, describing preliminary results. In \citep{9310895}, the details of the system orchestration were explained, and the authors proposed the deployment of containerized virtualized 5G infrastructure orchestrated by Kubernetes and Mosaic 5G Operator, with broad slicing radio support. In \citep{Nvoa2020RANSU}, the authors focused on a Cloud RAN (C-RAN) architecture implementation using OAI and a Software Defined RAN (SD-RAN) implementation, including FlexRAN (the first version of OAI's RIC) for RAN slicing. Network elements were virtualized using Docker container implementation and orchestrated with Kubernetes. RAN slicing on the FlexRAN platform and its API communication were demonstrated in an orchestrated and containerized scenario. In \citep{9318327}, a new OAI-based prototyping tool was used with a new OAI stub interface located at the MAC layer, capable of emulating all the underlying software and hardware of LTE V2X technology. The proposed solution delivers packets to target UE according to statistical packet loss and delay models. In \citep{9154290}, simulation modeling of the OpenAirInterface (OAI) C-RAN test environment deployed at Eurecom was proposed, and the performance evaluation of a Fast Calibration (FC) scheme based on antenna clustering was presented. The ray tracing technique generated an indoor line-of-sight (LOS) radio wave propagation model.

In 2021, several works focused on performance tests. In \citep{9633433}, the transmission performance measurement of a real 5G SA communication system was carried out, completing the integration and connection test of the OAI platform and the universal software radio peripheral (USRP). In \citep{9739187}, some key parameters were varied to analyze the performance of a 5G network. They analyzed parameters such as resource utilization in time, frequency, modulation and coding scheme (MCS), and transmission and reception gains. The performance of the OAI 5G New Radio (NR) was evaluated, and its dependence on parameters was quantified using three different connection models. In \citep{9442028}, the authors described a test platform built in South Africa for experimental work on mobile networks. The test platform uses the OAI RAN, a Fraunhofer core network, and the USRP X310. In \citep{9610039}, a Low-Density Parity Check (LDPC) accelerator based on Field-programmable Gate Arrays (FPGA) fully integrated into a software platform was proposed, capable of achieving a decoding throughput of more than 1.6~Gbps and up to 13x faster execution times compared to single-core software implementations.

In 2022, we can also see a lot of work on performance tests but also included resource consumption analyses. In \citep{9771860}, we seek to demystify the capabilities of existing tools and present guidelines that facilitate the improvement and development of additional features, addressing resource scaling and providing an analysis of the code and instructions available to facilitate the development of new algorithms. The authors discussed the scalability implications of network slicing and virtual network functions (VNF) of the 5G core. Containerized deployment of a network with the OAI 5GC and the gNBSIM RAN simulator~\citep{9771860} was demonstrated in a Kubernetes environment. An environment with abundant computational resources was used to deploy a large-scale deployment on OpenStack. In \citep{9880817}, authors present an analysis of arrival direction algorithms for application in cellular positioning technologies. The authors proposed an error-reducing positioning method and evaluated the performance in a 5G-NR system based on OAI in two scenarios with different multipath propagation conditions. The proposed method improves positioning performance and achieves an accuracy that meets the requirements of commercial use cases specified by 3GPP~\citep{3gpp.38.855}. In \citep{9771597}, tutorial material about resource scheduling in OAI-based systems is presented. A performance evaluation of up to 10 UEs was demonstrated, comparing the proposed and legacy scheduling algorithms. In \citep{10028998}, authors deployed an OAI-based 5G system to make video calls and download video streaming. Some use cases demonstrated video transmission in high quality and high resolution.

In 2023, performance tests are focused on different scenarios and applications. In \citep{10170129}, similar to \citep{9771597}, authors compare some scheduling algorithms using the OAI system. As expected, the throughput performance using Round Robin is more uniform among UEs compared to resource allocation based on FIFO fashion. In \citep{10118776}, authors investigated system coverage parameters, including the Reference Signal Received Power (RSRP) and Reference Signal Received Quality (RSRQ), and the signal-to-interference and noise ratio (SINR) for both single-user and multi-user scenarios. They also evaluated downlink and uplink data rate and latency for deployment with 60~MHz bandwidth. The data rate reached up to 390 Mbps, and the average latency was 19~ms, while the minimum latency value was 6~ms. In \citep{10041332}, a new 5G MEC-OAI test platform was presented, adhering to existing standards. It leverages standards while keeps the flexibility of virtualized infrastructure. A video streaming application was demonstrated, addressing object detection scenarios.

Although all the works mentioned above contribute to adopting OAI for research and demonstration initiatives, they lack tutorial material for reproducibility. This paper provides such material, configuration discussions, and a throughput benchmarking evaluation of the 5G OAI using USRP.

\section{OpenAirInterface (OAI)}
\label{sec_virtual}

Founded in 2014 by EURECOM, the OpenAirInterface Software Alliance (OSA) is a non-profit organization based in France that acts as the operation center for OpenAirInterface (OAI) software, being responsible for the development of its roadmap, quality control, and dissemination of its solutions \citep{oai-about}. The OAI project is designed to provide complete end-to-end 4G LTE and 5G NR systems, compliant with 3GPP vital elements, such as the RAN (monolithic or split mode), CN and UE~\citep{KALTENBERGER2020107284, oai-about}. 

OAI is also an open-source software, allowing researchers and developers to implement new patented technologies without waiving their private property rights. In addition, due to the possibility of using SDR platforms, it has become feasible to operate a 5G network with the transmission of electromagnetic signals. As the OAI code implements real signaling, it is possible to use commercial devices, such as smartphones, to carry out end-to-end connectivity experiments. This makes the software widely attractive in academia and industry, and the only requirement for such an experiment is to have SDR hardware compatible with the OAI code ~\citep{KALTENBERGER2020107284}.

OAI has an international community of developers and is organized into three main projects:

\begin{itemize}
    \item \textbf{OAI 5G RAN}: responsible for developing the 4G and 5G protocol stacks as specified in 3GPP~\citep{KALTENBERGER2020107284}. In its roadmap, one can find the implementation of essential features like Cell-Free MIMO, FR2 operation, sidelink for V2X, Xn handover procedures, and 5G RAN slicing;
    \item \textbf{OAI 5G Core Network (CN)}: responsible for developing 4G (EPC) and 5G (5GC) network cores according to 3GPP specifications~\citep{oai-cn,KALTENBERGER2020107284}. Such a group is responsible for implementing essential features like support for Traffic Steering redirection, QoS support for Control and Data Planes, and configuration of 5GC with REST API;
    \item \textbf{OAI Operations and Maintenance (OAM)}: responsible for implementing and maintaining different management interfaces proposed by O-RAN Alliance~\citep{oai-oam}. It aims to develop essential features related to non-RT, near-RT RIC, and O-RAN open interfaces. 
    
\end{itemize}
All projects are discussed in the following sections.

\subsection{OAI 5G RAN}

The OAI 5G RAN project handles a RAN solution compliant with 3GPP and implements both SA and NSA architectures, \citep{oai-ran}. Moreover, the OAI SA architecture supports two modes of operation, namely, split and monolithic modes:

\begin{itemize}
    \item The \textbf{split mode} divides the RAN into three components: Next Generation NodeB Centralized Unit (gNB-CU), Next Generation NodeB Distributed Unit (gNB-DU), and Radio Unit (RU). The gNB-CU is responsible for the Radio Resource Control (RRC) and Packet Data Convergence Protocol (PDCP) layers, while the gNB-DU is responsible for the Physical (PHY), Media Access Control (MAC), and Radio Link Control (RLC) layers. The RU is responsible for the entire radio part of the RAN \citep{KALTENBERGER2020107284};
    \item In \textbf{monolithic mode}, the different components of the RAN are part of a single application, without separation \citep{KALTENBERGER2020107284}.
\end{itemize}

In this paper, we used the SA architecture in the monolithic mode due to its better stability in the specific OAI code version utilized.

\subsection{OAI Operations and Maintenance (OAM)}

The OAI OAM project focuses on orchestration, monitoring, and maintenance of OAI RAN and Core Network Functions. Furthermore, it implements and maintains the interfaces proposed by the O-RAN Alliance. The project offers solutions such as the FlexRIC, an O-RAN-compliant RIC, an E2 interface, and a Multi-access Edge Computing Platform (MEP) \citep{oai-oam}.

\subsection{OAI 5G Core Network}

The OAI 5G CN is a 3GPP-compliant project that focuses on the Network Functions (NFs) implementation of both 5G Standalone (SA) and 5G Non-standalone (NSA) \citep{oai-cn}. In addition, the OAI 5G CN can be adapted to support different scenarios, in which its deployment can be done in three ways:

\begin{itemize}
\item \textbf{Minimalist 5GC}: The most basic version, comprising the virtual network functions Access and Mobility Management Function (AMF), Session Management Function (SMF), Network Repository Function (NRF) and User Plane Function (UPF) \citep{oai-cn};

\item \textbf{Basic 5GC}: Version with the same components as Minimalist mode plus the Unified Data Management (UDM), Authentication Server Management Function (AUSF) and Unified Data Repository (UDR) \citep{oai-cn};

\item \textbf{Slicing 5GC}: The most complete mode, containing all the Basic 5GC functions plus the Network Slicing Selection Function (NSSF) \citep{oai-cn}.

\end{itemize}

Moreover, about UPF, the OAI provides three flavors: SPGW-U-tiny, inherited from 4G but with the addition of some 5G features; VPP-UPF, based on Vector Packet Processing (VPP), offering high performance, modularity, and flexibility; and Production grade UPF, which is still under development and will use SD-Fabric \citep{oai-cn, upf-vpp-git, upf-vpp-oai-git}.

Finally, OAI has three deployment platforms: bare-metal installation or using virtual machines; automated deployment in Docker containers, using Docker-Compose; and Cloud-native using Helm Chart~\citep{oai-cn}.

\begin{figure}[h!]
  \centering
  \includegraphics[width=\linewidth]{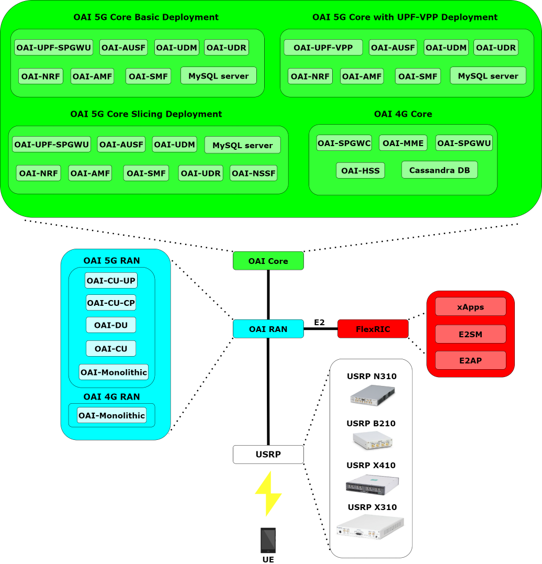}
  \caption{OAI deployment options.}
  \label{fig:oai_general}
\end{figure}

Figure~\ref{fig:oai_general} illustrates all deployment options of 4G and 5G networks using OAI.

\section{Installation/Deployment Tutorial}
\label{sec_install}


In this paper, a containerized approach is adopted for the sake of installing and deploying the OAI stack in bare metal. The minimum requirements for the OAI installation are a desktop/server with 4 CPU cores, 16 GB of RAM, and a minimum of 1.5 GB of free storage space for Docker images. Ubuntu 20.04 LTS or newer is also needed, along with Python 3, Docker, and Docker Compose. We provide two shell scripts so that automating the deployment of the 5G core and gNB, including all dependencies and configurations. All containers run in privileged mode.

The installation process starts by downloading the repository \url{https://github.com/lance-ufrn/oai_bench_tuto} and executing the two available scripts:

\begin{enumerate}
    \item Firstly, the 5GC script (\textit{oai\_5g\_core\_install.sh}) should be executed. It installs all necessary dependencies and
    certificates to activate the 5GC, deploying it at the end;
    \item After that, the script \textit{services.sh} should be called to install and deploy the gNB. This script could receive input parameters for different gNB configurations. 
\end{enumerate}

Our repository has some configuration files, the OAI 5GC files, and two scripts mentioned above to deploy 5GC and gNB. The script named \textit{oai\_5g\_core\_install.sh} is responsible for:

\begin{itemize}
    \item installing the essential dependencies such as ca-certificates, curl, gnupg, and lsb-release;
    \item installing Wireshark, an optional tool useful for packet analysis;
    \item configuring Ubuntu to allow IP packet forwarding between network interfaces and modifying the \textit{iptables} filtering table to accept the forwarding. 
    \item executing Docker-related operations, such as: (i) removing any Docker-related files from the system's package source list; (ii) creating the suitable directory to store package repository keys; downloading the Docker GPG key and decrypting it; (iii) adding a Docker repository to the package sources list file, specifying the system architecture, GPG key to be used, and the stable version of Docker; (iv) installing Docker Engine (docker-ce), Docker Command Line Interface (docker-ce-cli), contained (containerd.io), Docker Compose plugin, and the Docker Compose binary version 1.29.2 corresponding to the system architecture; 
    \item stopping and removing all running containers, as well as removing all locally present images and downloading specific Docker Hub images;
    \item tagging the downloaded images are tagged with specific tags, which facilitate referencing during the container execution process;
    \item executing the synchronization of 5GC components;
    \item Lastly, executing the Python command responsible for starting the basic 5GC.
\end{itemize}

After the installation process, 5GC images become visible, namely the AMF (Access and Mobility Management Function), NRF (Network Registration Function), SMF (Session Management Function), UDR (User Data Repository), UDM (Unified Data Management), AUSF (Authentication Server Function), and SPGW-U Tiny (Packet Gateway - User Plane Functionality), as well as the TRF (Traffic Generator).

Before proceeding with the gNB script, we must ensure that the USRP (e.g., N310 or B210) is connected to the host machine by verifying its IP address. Once USRP is properly connected to the host machine, we must execute \textit{./services.sh [options]}, where \textit{[options]} is one of the listed options as follows:

\begin{itemize}
    \item \textit{n106} for 106 Physical Resource Blocks (PRBs) of bandwidth into a USRP N310;
    \item \textit{n162} for 162 PRBs of bandwidth into a USRP N310;
    \item \textit{n273} for 273 PRBs of bandwidth into a USRP N310;
    \item \textit{b106} for 106 PRBs of bandwidth into a USRP B210;
    \item \textit{b162} for 162 PRBs of bandwidth into a USRP B210;
    \item \textit{stop} to finalize the gNB containers.
\end{itemize}

For instance, the command \textit{./services.sh n106} installs and deploys a gNB with 162 PRBs into a USRP N310. After executing this command, the script \textit{./services.sh} is in charge of building the Docker container entailing a gNB image with the specified bandwidth (number of PRBs) and the chosen USRP.

In order to verify if the gNB is correctly connected to the 5GC, we could analyze the logs of the oai-amf container using the command \textit{docker logs oai-amf -f}. The AMF logs indicate the connection status of 5GC, gNB, and its connected UEs.

\section{Test Scenarios}
\label{sec_TestScen}

Our tests focus on throughput analyses by varying the system bandwidth of a carrier starting from 3.7~GHz. The over-the-air experiments use a USRP N310 to implement the RF front-end and the Low-PHY layer following the 7.2x split option. A Motorola G50 is used as UE. Table~\ref{tab:tests-summary} summarizes bandwidth test scenarios.

\begin{table}[h!]
    \centering
    \caption{Test scenarios and bandwidths.}
    \begin{tabular}{p{1cm}p{3cm}p{1.7cm}}
    \hline
        PRB & Band Identification & Bandwidth     \\\hline
        106 & N78                 &               40MHz\\
        162 & N78                 &               60MHz\\
        273 & N78                 &               100MHz\\\hline
    \end{tabular}
    \label{tab:tests-summary}
\end{table}

\subsection{Infrastructure Description}
\label{sec_infra_hardware_software}

We use a Lenovo server, model ThinkSystem SR630, equipped with the Intel Xeon Gold 5218R processor, which has 20 Cores and 40 threads at 2.10GHz, 128 GB of RAM, and a 10TB HDD. The Operating System (OS) is Ubuntu Desktop 20.04.1 LTS, Kernel 5.15.0-89, Docker version 24.0.7, and Docker Compose version 2.21.0.

\begin{figure*}[h!]
  \centering
  \includegraphics[width=\linewidth]{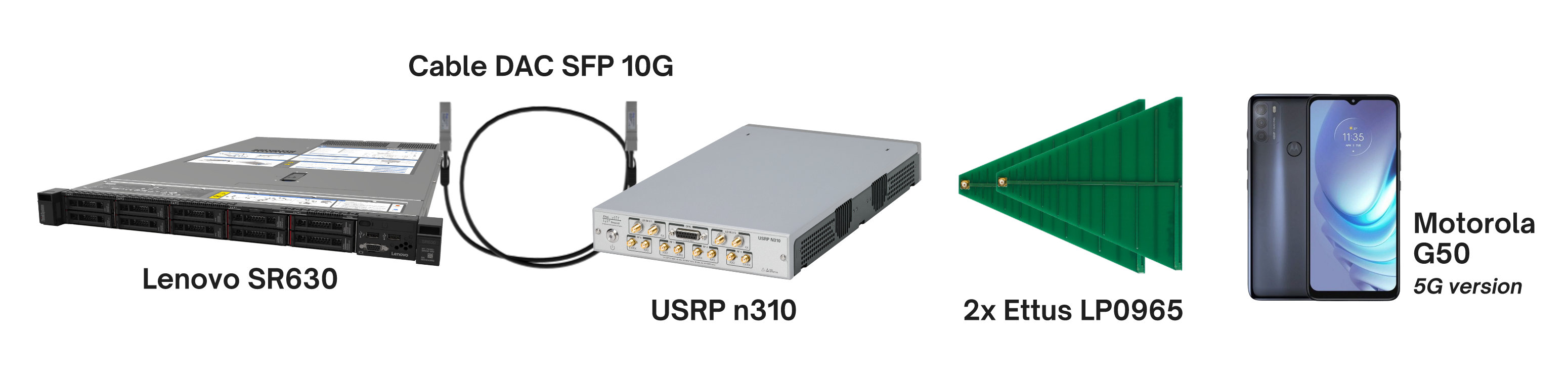}
  \caption{Infrastructure Hardware Resources.}
  \label{fig:infrastructure-resources}
\end{figure*}

The RF front-end uses two Ettus LP0965 antennas. The USRP N310 is directly connected to the server by a Direct Attach Copper (DAC) cable with 10Gbps SFP+ communication interfaces. Figure \ref{fig:infrastructure-resources} shows the physical infrastructure used in the tests.

All tests use the Motorola G50 5G version equipped with the MediaTek Dimensity 700 (MT6833) chipset~\citep{devicespecifications-moto-g50-specification}. It is compatible with 5G FR1 bands, including N77 and N78 for Stand Alone (SA) and Non-Stand Alone (NSA) networks~\citep{phonemore-moto-g50-specification}. The installed OS is Android version 11/12, and our tests are executed using the phone's own battery. The battery charge is always higher than 50\% during the tests without connection to another power source. The temperature in the test environment ranged between 21ºC and 25ºC.

\subsection{gNB Parameters configuration}
\label{ParametersConfiguration}

This section details the configuration of the parameters to achieve the highest possible stability and throughput in the multiple data transfer tests between UE and 5GC.

Part of the gNB's \textit{.conf} file are in Listings \ref{code:PhysicalParameters} and \ref{code:SISOconfigurarion} for SISO mode operation. The parameter \texttt{min\_rxtxtime} is related to the minimum duration for which a receiver must be active to process incoming signals or a transmitter must be active to send out signals, we set it in 5 ms. One could disable 256-QAM by setting \texttt{force\_256qam\_off} to 1. The parameters \texttt{do\_CSIRS} and \texttt{do\_SRS} activate MIMO and should be disabled by assigning them 0. Lastly, the antenna ports must be configured as 1 for SISO operation.

\begin{lstlisting}[label=code:PhysicalParameters,caption=SISO configuration: Part 1.]
gNBs = ({
  do_CSIRS              = 1; 
  min_rxtxtime          = 5;
  do_SRS                = 0; 
  force_256qam_off      = 0;
  pdsch_AntennaPorts_N1 = 1;
  pdsch_AntennaPorts_XP = 1;
  pusch_AntennaPorts    = 1;
)}
\end{lstlisting}

Regarding Listing \ref{code:SISOconfigurarion}, \texttt{nb\_tx} and \texttt{nb\_rx} also determine the operation mode as either SISO or MIMO. They should be set to 1 for SISO and 2 or 4 for MIMO. Both \texttt{att\_tx} and \texttt{att\_rx} introduce attenuation in the transmitter, ranging from 6 to 10. They are configured as 10 for maximum attenuation and 0 for no attenuation. The parameter \texttt{bands} selects the operating frequency following the 3GPP nomenclature, and in our case, we selected 78 for n78 band~\citep{3gpp_38.101-1}. The parameter \texttt{max\_pdschReferenceSignalPower} adjusts the power of the PDSCH reference signal, measured in dBm. For the \texttt{max\_rxgain}, which controls reception gain, the recommended value for the N310 is 75~\citep{bubble-max-gain-rx}.

\begin{lstlisting}[label=code:SISOconfigurarion,caption=SISO configuration: Part 2.]
RUs = (
{
  local_rf       = "yes";
  nb_tx          = 1;
  nb_rx          = 1;
  att_tx         = 0;
  att_rx         = 0;
  bands          = [78];
  max_pdschReferenceSignalPower = -27;
  max_rxgain                    =  75;
  eNB_instances  = [0];
  bf_,weights=[0x00007fff, 0x0000,
                0x0000, 0x0000];
  sdr_addrs = "addr=192.168.20.2,
                clock_source=internal,
                time_source=internal"
}
\end{lstlisting}

Listing 4 presents the parameters that adjust the time slots. While \texttt{dl\_UL\_TransmissionPeriodicity} defines the transmission periodicity of downlink and uplink slots, \texttt{nrofDownlinkSlots} defines the number of slots with only downlink symbols. The latter must be an integer between 0 and the maximum number of slots per frame. The parameter \texttt{nrofDownlinkSymbols} defines the number of symbols per downlink slot and must be an integer between 0 and the maximum number of symbols minus 1. Similarly, the parameter \texttt{nrofUplinkSlots} defines the number of slots with only uplink symbols, and \texttt{nrofUplinkSymbols} represents the number of symbols per uplink slot.

\begin{lstlisting}[label=code:SlotsConfigurations,caption=Slots Configurations.]
servingCellConfigCommon = ({
  dl_UL_TransmissionPeriodicity = 6;
  nrofDownlinkSlots             = 6;
  nrofDownlinkSymbols           = 7;
  nrofUplinkSlots               = 3;
  nrofUplinkSymbols             = 6;
  
  #Other parameters of servingCellConfigCommon
});
\end{lstlisting}

An essential set of parameters are the ones responsible for defining the frequency and bandwidth. Once these parameters are defined, not in terms of Hz, but in terms of Absolute Radio Frequency Channel Number (NR-ARFCN), the network planner has finalized the basic configuration of the experiment. The NR-ARFCN is defined on the 3GPP TS 38.104~\citep{3gpp_38.104}.

\begin{figure}[h!]
  \centering
  \includegraphics[width=\linewidth]{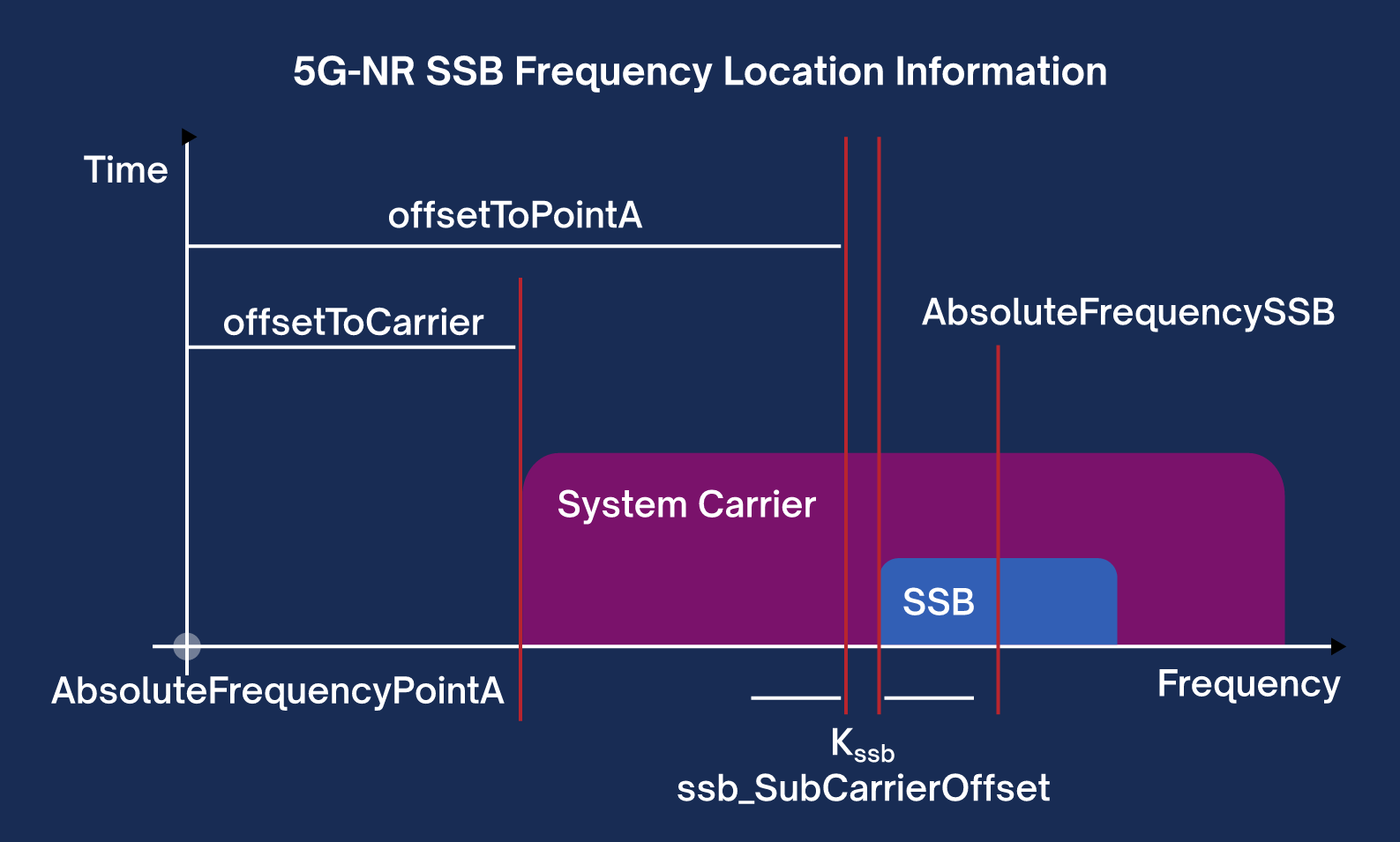}
  \caption{Infrastructure Resources.}
  \label{fig:diagram_5gnr}
\end{figure}

We should calculate and define the values of \texttt{dl\_absoluteFrequencySSB} and \texttt{dl\_absoluteFrequencyPointA}, as illustrated in Figure \ref{fig:diagram_5gnr}. 

Note that \texttt{absoluteFrequencySSB} is localized in the center of SSB, but it can be localized in the center of bandwidth, according to the network planner's decision. If \texttt{offsetToCarrier} is 0, the \texttt{absoluteFrequencyPointA} is localized at the beginning of the bandwidth. In our experiments, we have set the \texttt{absoluteFrequencySSB} to be at the center of bandwidth and \texttt{offsetToCarrier} is 0. Generally, both parameters are measured in Hz, but we need to get their corresponding values in NR-ARFCN to configure the gNB \textit{.conf} file. Equation~\ref{eq:nrarfcn1} shows how to convert the values from NR-ARFCN to Hz, while Equation~\ref{eq:nrarfcn2} from Hz to NR-ARFCN.

\begin{equation}
    \label{eq:nrarfcn1}
     F_{R} = F_{R\_Offs} + \Delta F_{G} (N_{R} - N_{R\_Offs}), 
\end{equation}

\begin{equation}
    \label{eq:nrarfcn2}
    N_{R} = N_{R\_Offs} + \frac{(F_{R} - F_{R\_Offs})}{\Delta F_{G} }, 
\end{equation}

\noindent where $F_R$ is the reference frequency in Hz, $\Delta F_G$ is the granularity of the global frequency raster, $F_{R\_Offs}$ is RF reference frequency offset, $N_R$ is the NR-ARFCN, and $N_{R\_Offs}$ is NR-ARFCN reference offset. Typical values of those parameters are shown in Table~\ref{tab:nrarfcn}. 

\begin{table}[ht]
\caption{NR-ARFCN Parameters.}
\label{tab:nrarfcn}
\begin{tabular}{cccc}
\hline
\textbf{\begin{tabular}[c]{@{}c@{}}Range of \\frequencies (MHz)\end{tabular}} &  \textbf{\begin{tabular}[c]{@{}c@{}}$\Delta F_{G}$ \\(kHz)\end{tabular}}  & \textbf{\begin{tabular}[c]{@{}c@{}}$F_{R\_Offs}$ \\(MHz)\end{tabular}} & \textbf{$N_{R\_Offs}$} \\ \hline
0-3000                                                                        & 5                            & 0                          & 0                    \\
3000-24250                                                                    & 15                           & 3000                       & 600000               \\
24250-100000                                                                  & 60                           & 24250.08                   & 2016667              \\ \hline
\end{tabular}
\vspace{1cm}
{\footnotesize \textbf{Source:} Reproduction of Table 5.4.2.1-1 of 3GPP TS 38.104.} 
\end{table}

For example, if the \texttt{absoluteFrequencyPointA} is $3.7$ GHz with $106$ PRBs, SCS $30$ kHz and $12$ OFDM symbols, then the total bandwidth would be $106 \cdot 30 \cdot 12 = 38.16 MHz$. Therefore the \texttt{absoluteFrequencySSB} would be 3719.08~MHz ($3700 + \frac{38.16}{2} $ = 3719.08 MHz).  However, a rounding is needed because all frequency values must be multiples of 1.44~MHz, the smallest bandwidth of LTE systems. It is safe to round the \texttt{absoluteFrequencySSB} to $3720$~MHz and adjust \texttt{absoluteFrequencyPointA} to $3720 - 38.16/2 = 3702.92$ MHz to maintain the total bandwidth as previously calculated in $38.16$ MHz. Then, the network planner can use the Equation \ref{eq:nrarfcn2} to convert the new values found of \texttt{absoluteFrequencyPointA} and \texttt{absoluteFrequencySSB} in Hz to NR-ARFCN.

Thus, to calculate the new value of \texttt{absoluteFrequencySSB} (3720 MHz), from Equation \ref{eq:nrarfcn2}, values are in MHz:

\begin{equation}
    \label{eq:nrarfcn3}
    N_{R_{SSB}} = 600000 + \frac{(3720 - 3000)}{0.015} = 648000.
\end{equation}

Next, one way to calculate \texttt{absoluteFrequencyPointA} parameter is considering the value of \texttt{absoluteFrequencySSB} raster, previously found in Equation \ref{eq:nrarfcn3}. This value corresponds to the center bandwidth in the frequency raster. In other words, the network planner only needs to find the value corresponding to half of the bandwidth in raster and then remove it from the result found in Equation \ref{eq:nrarfcn3} to find \texttt{absoluteFrequencyPointA} in raster.

Hence, because we are calculating a reference value in the frequency range between 3 GHz and 24.25 GHz, $N_{R\_Offs}$ can be considered 0, from Equation \ref{eq:nrarfcn2}:

\begin{equation}
    \label{eq:nrarfcn4}
    N_{R_{RefBW}} = 0 + \frac{(3719.08 - 3000)}{0.015} = 1272.
\end{equation}

Then, \texttt{absoluteFrequencyPointA} is:

\begin{equation}
    \begin{split}
    \label{eq:nrarfcn5}
    N_{R_{PointA}} &=  N_{R_{SSB}} - N_{R_{RefBW}} \\ &= 648000 - 1272 = 646728.
    \end{split}
\end{equation}

The parameters used in the experiments with 106 PBRs are shown in Listing~\ref{code:frequencyConfigurations106PRB}. The \texttt{dl\_frequencyBand} parameter defines the possible frequency bands and bandwidths of the 5G NR channel. We chose 78 for the n78 frequency band. The parameter \texttt{dl\_absoluteFrequencyPointA} points to the first resource element of the 0-th resource block of the common resource block grid. That resource element, which is the initial frequency of the band, is 3.7 GHz for this work. The parameter \texttt{absoluteFrequencySSB} represents the center frequency of the SSB. It is set as 3.72 GHz because we have 40 MHz of bandwidth. The \texttt{dl\_carrierBandwidth} parameter defines the carrier bandwidth in number of PRBs. For 106 PRBs, we have 40~MHz of bandwidth. The \texttt{initialDLBWPlocationAndBandwidth} parameter defines the carrier location and bandwidth. The parameters \texttt{ul\_carrierBandwidth} and \texttt{initialULBWPlocationAndBandwidth} must match the parameters \texttt{dl\_carrierBandwidth} and \texttt{initialDLBWPlocationAndBandwidth}.

\begin{lstlisting}[label=code:frequencyConfigurations106PRB,caption=Frequency Configurations for 106 PRB.]
servingCellConfigCommon = ({
  dl_frequencyBand                 = 78;
  absoluteFrequencySSB             = 648000;
  dl_absoluteFrequencyPointA       = 646728;
  dl_carrierBandwidth              = 106;
  initialDLBWPlocationAndBandwidth = 28875;
  ul_carrierBandwidth              = 106;
  initialULBWPlocationAndBandwidth = 28875;
  
  #Other parameters of servingCellConfigCommon
});
\end{lstlisting}

Listening \ref{code:frequencyConfigurations162PRB} and \ref{code:frequencyConfigurations273PRB} show the configurations for 60 and 100~MHz, i.e., for 162 and 273 PRBs, respectively. The rest of the parameters are changed accordingly.

\begin{lstlisting}[label=code:frequencyConfigurations162PRB,caption=Frequency Configurations for 162 PRB.]
servingCellConfigCommon = ({
  dl_frequencyBand                 = 78;
  absoluteFrequencySSB             = 648672;
  dl_absoluteFrequencyPointA       = 646728;
  dl_carrierBandwidth              = 162;
  initialDLBWPlocationAndBandwidth = 31624;
  ul_carrierBandwidth              = 162;
  initialULBWPlocationAndBandwidth = 31624;
  
  #Other parameters of servingCellConfigCommon
});
\end{lstlisting}

\begin{lstlisting}[label=code:frequencyConfigurations273PRB,caption=Frequency Configurations for 273 PRB.]
servingCellConfigCommon = ({
  dl_frequencyBand                 = 78;
  absoluteFrequencySSB             = 650016;
  dl_absoluteFrequencyPointA       = 646728;
  dl_carrierBandwidth              = 273;
  initialDLBWPlocationAndBandwidth = 1099;
  ul_carrierBandwidth              = 273;
  initialULBWPlocationAndBandwidth = 1099;
  
  #Other parameters of servingCellConfigCommon
});
\end{lstlisting}

\subsection{UE configuration}
\label{sec_case1}


In order to facilitate the installation and configuration of the SIM card, we developed an illustrated step-by-step guide, available at \url{https://github.com/lance-ufrn/oai_bench_tuto} as open source. This guide assists with the installation and configuration procedures of UEs. The tutorial entails the establishment and configuration process of a development environment for Sysmocom SIM cards on the Linux operating system, as well as the necessary steps to be taken on the UEs to enable their usage.

It is possible to use SIM cards from other manufacturers, but it is necessary to configure the parameters used in this tutorial: \textbf{International Mobile Subscriber Identity(IMSI)} is a number that uniquely identifies all UEs in a cellular network; \textbf{Derived Operator Code (OPc)} is a key used in UE authentication algorithms; and \textbf{Authentication Key (Ki)} that is a key used with the OPc for UE authentication.

The values of these parameters must be compatible with the 5G network configured. The range of values for IMSI and the fixed values for OPc and Ki that are compatible with this tutorial are shown below:

\begin{itemize}
    \item \textbf{IMSI}: 001010000000001 to 001010000000003;
    \item \textbf{OPc}: c42449363bbad02b66d16bc975d77cc1; 
    \item \textbf{Ki}: fec86ba6eb707ed08905757b1bb44b8f.
\end{itemize}

\section{Results}
\label{sec_Results}

This section puts forward presentation and discussions about three use cases harnessing our OAI testbed. The first use case, presented in Section~\ref{sec_sp_ocup}, analyses the spectrum occupancy for a 5G gNB with 40, 60, and 100~MHz bandwidth. We show three spectrograms, a visual representation of the spectrum of frequencies of a signal as it varies with time. The second use case, in Section~\ref{sec_cov_perf}, provides wireless coverage evaluation by means of the  Radio Environment Map (REM) of RSRP. The third use case, in Section~\ref{sec_tput_perf}, comes out throughput and Round Trip Time (RTT) performance when varying the total number of PRBs, as well as for a UE-to-UE connection and for a varying distance.

\begin{figure}[h!]
    \centering
    \includegraphics[width=\linewidth]{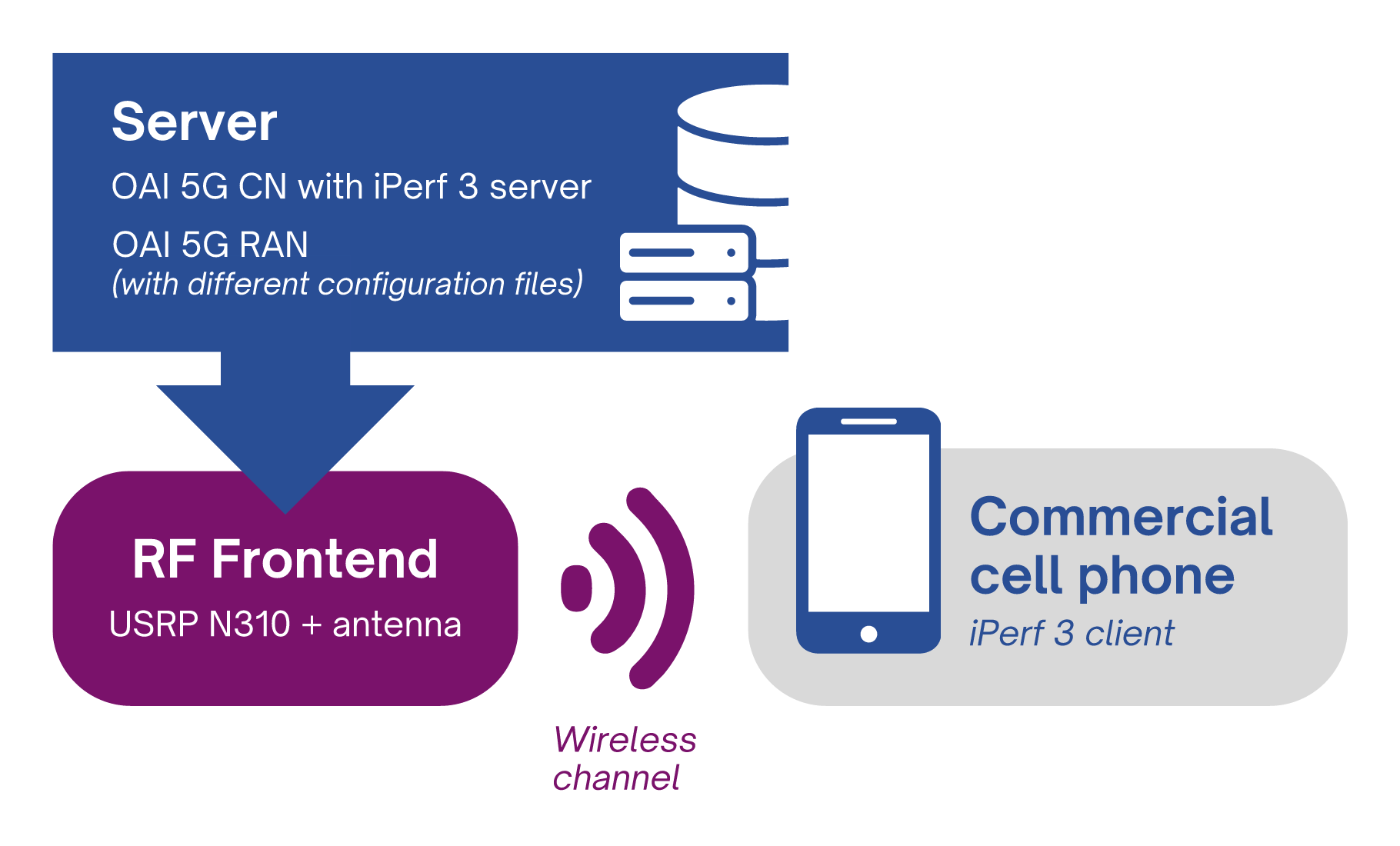}
    \vspace*{2mm}
    \caption{Testbed high-level view for use case experiments.}
    \label{fig:fig03}
\end{figure}

Figure~\ref{fig:fig03} depicts a high-level view of our testbed, for running the above mentioned use cases atop the equipments described in Section~\ref{sec_infra_hardware_software}.

We attach a UE to the gNB for both use cases to measure spectral occupancy and throughput using iPerf3, which is built on a client-server model that comes up with maximum Transmission Control Protocol (TCP) throughput measurements between client and server endpoints. Packet loss, RTT, and Jitter using the well-known ping command are also evaluated. In our testbed, the client is installed on the UE side while the server is in the 5GC. We run iPerf3 for 180 seconds and perform measurements for the PRB values of 106, 162, and 273, and distances from 1 to 12~meters.

\subsection{Use Case 1: Spectrum Occupation}
\label{sec_sp_ocup}

\begin{figure}[h!]
    \centering
    \begin{subfigure}[b]{\columnwidth}
        \centering
        \includegraphics[width=0.95\textwidth]{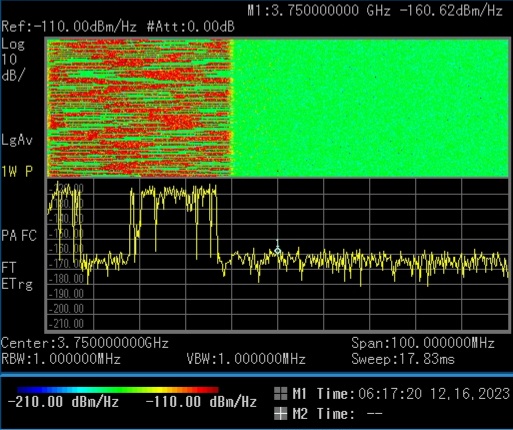}
        \caption{Bandwidth of 40 MHz.}
        \label{fig:fig2a}
    \end{subfigure}
    \hfill
    \begin{subfigure}[b]{\columnwidth}
        \centering
        \includegraphics[width=0.95\textwidth]{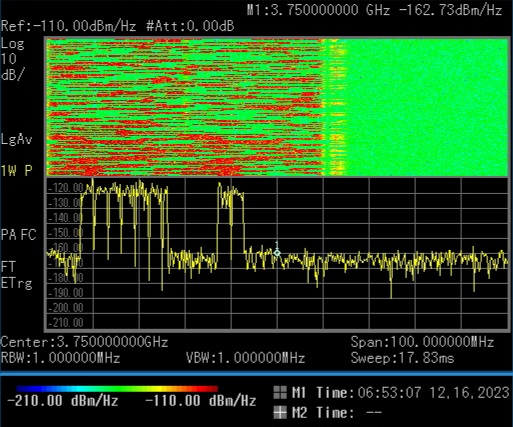}
        \caption{Bandwidth of 60 MHz.}
        \label{fig:fig2b}
    \end{subfigure}
    \hfill
    \begin{subfigure}[b]{\columnwidth}
        \centering
        \includegraphics[width=0.95\textwidth]{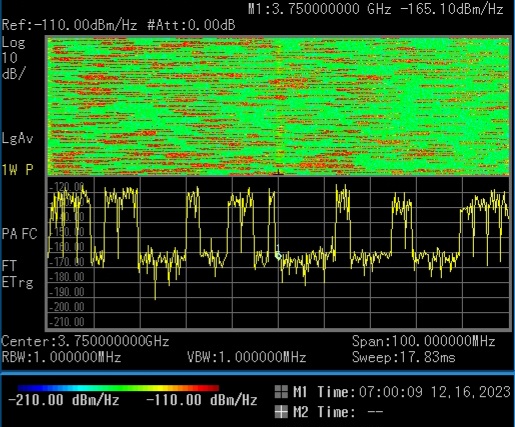}
        \caption{Bandwidth of 100 MHz.}
        \label{fig:fig2c}
    \end{subfigure}
    \caption{Spectrum occupation behavior for the bandwidth settings.}
    \label{fig:OccupiedSpectrum}
\end{figure}

In order to monitor the spectrum occupation behavior over time in the OAI testbed, the Keysight N9342C portable spectrum analyzer (HSA)\footnote{http://www.keysight.com/us/en/assets/7018-02496/data-sheets} was used. The \textit{.conf} files of the gNBs are configured to 40, 60, and 100 MHz bandwidths. Figure \ref{fig:OccupiedSpectrum} depicts the spectrum analyzer outputs for the three bandwidth settings. The spectrum trace displays a central frequency of 3.75 GHz, with a bandwidth of 100 MHz. The measured range extends from 3.7 GHz to 3.8 GHz, with x-axis plot gradations of 10 MHz.


\FloatBarrier

Figure \ref{fig:fig2a} illustrates the spectrogram and spectrum trace for 40 MHz of the occupied band. The spectrum is fully occupied during the transmission, and the spectrogram shows the 40~MHz bandwidth occupied. For these results, the gNB \textit{.conf} file is configured with 106 PRBs. Figure~\ref{fig:fig2b} illustrates the spectrogram and the spectrum trace for 60~MHz bandwidth. During the transmission, the spectrum is fully occupied, and the spectrogram shows the subcarrier's power spread out over 60~MHz. The gNB \textit{.conf} file is configured with 162~PRBs. Finally, Figure~\ref{fig:fig2c} illustrates the spectrogram and spectrum trace for 100 MHz configuration. For this last experiment, the gNB \textit{.conf} file is configured with 273 PRBs, and the whole 100~MHz has subcarrier power.

\subsection{Use Case 2: Coverage Evaluation}
\label{sec_cov_perf}

The wireless coverage within our laboratory has been analyzed by collecting the RSRP across different areas, providing a detailed understanding of signal strength and coverage patterns.

\begin{figure}[h!]
    \centering
    \includegraphics[width=\linewidth]{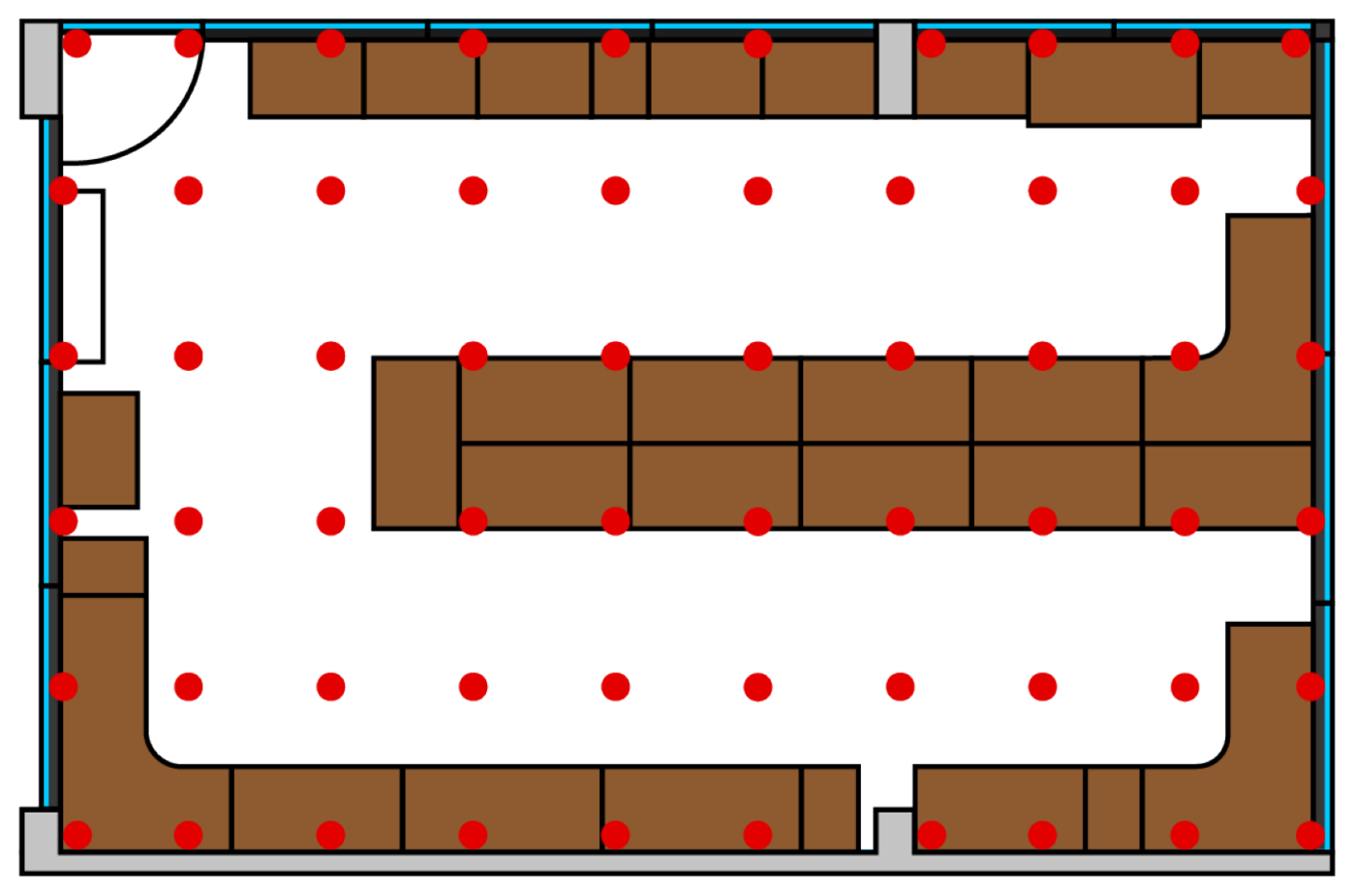}
    \caption{Illustration of the distribution of measurement points in the laboratory.}
    \label{fig_covMapMet}
\end{figure}

As illustrated in Figure~\ref{fig_covMapMet}, the laboratory layout features strategically placed measurement points (indicated by red dots) distributed throughout the space. The points are arranged in a grid pattern to ensure uniform coverage and accurate representation of the signal strength in various locations. An in-house-built Python script is employed to map and visualize the collected data, transforming raw measurements into a REM.

\begin{figure}[h!]
    \centering
    \includegraphics[width=\linewidth]{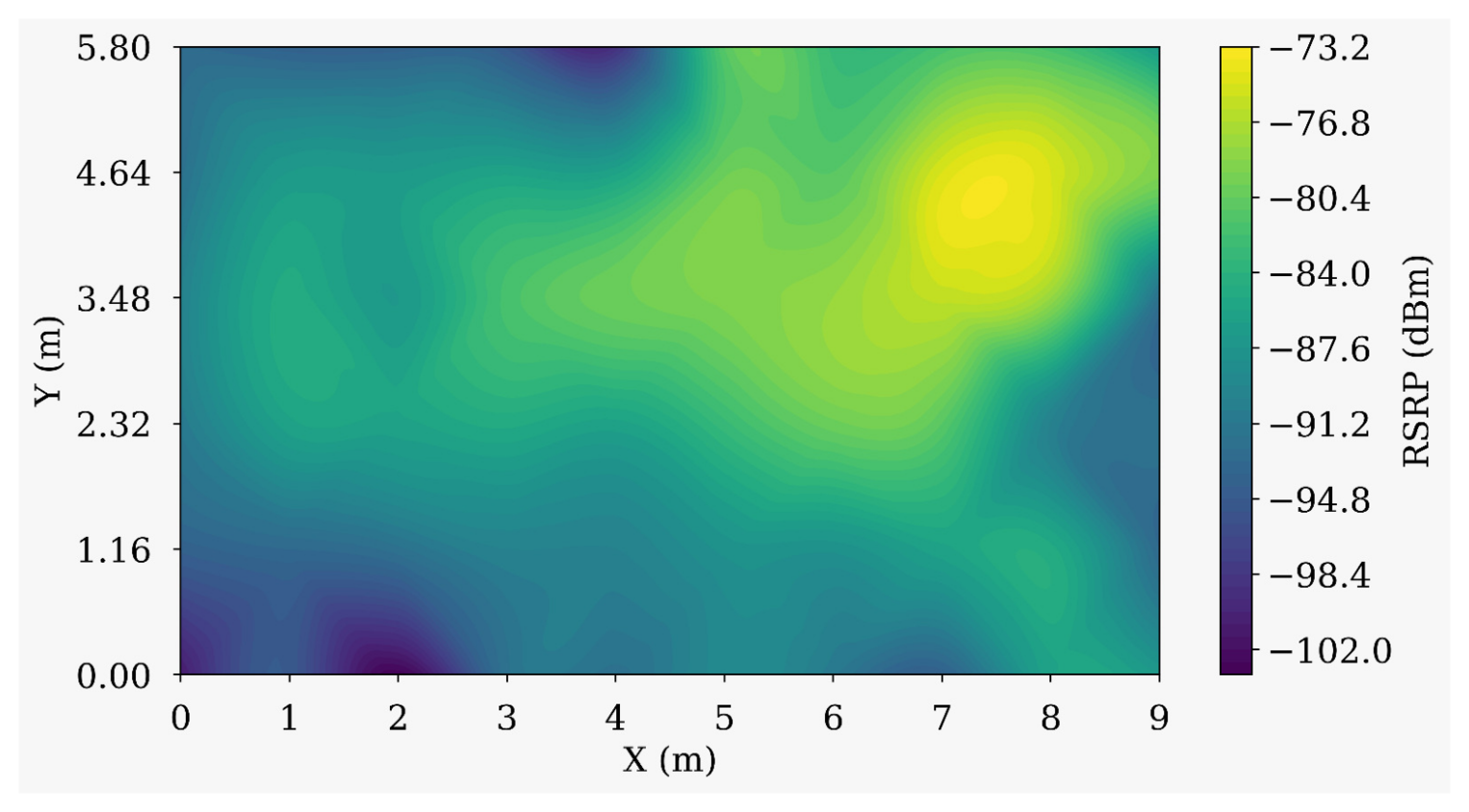}
    \caption{Measured Coverage Map.}
    \label{fig_covMap}
\end{figure}

The coverage map in Figure~\ref{fig_covMap} illustrates the RSRP measurements in our laboratory. The RSRP values, expressed in dBm, indicate the power level of the received signal across different locations within the lab. The color gradient on the map represents different power levels, with purple indicating the weakest signals and yellow indicating the strongest. The RSRP values range from -102.0 dBm to -73.2 dBm. The brightest yellow region, indicating the highest signal strength, is located near the transmission setup. The green to light green regions, covering the central part of the lab, show RSRP values ranging from -84.0 dBm to -76.8 dBm. These areas have moderately strong signals, which should be sufficient for most wireless communication needs. The blue and purple regions, particularly towards the bottom left corner, indicate the weakest signal levels, with RSRP values ranging from -94.8 dBm to -102.0 dBm. These areas are farthest from the signal source and may experience connectivity issues or lower data rates. Thus, our laboratory's coverage map highlights significant signal strength variations across different areas. While most of the lab enjoys adequate to strong signal levels, some regions, particularly towards the lower left, experience weaker signals.

\subsection{Use Case 3: Benchmarking Tests}
\label{sec_tput_perf}

\subsubsection{Channel-Ideal Scenario}
\label{sec_ideal}

For the throughput benchmarking tests, we perform 10~transmissions of 180 seconds for each PRB value to improve the confidence interval.

\begin{figure}[h!]
    \centering
    \includegraphics[width=1\linewidth]{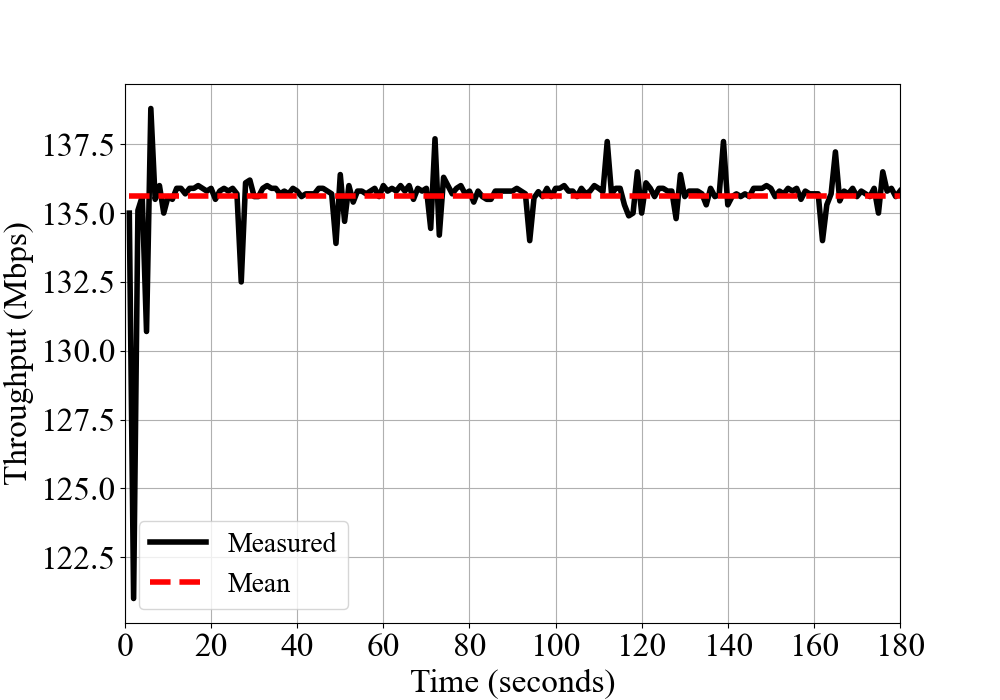}
    \caption{Throughput measurement and mean for 106 PRBs.}
    \label{fig:fig04}
\end{figure}

Figure \ref{fig:fig04} illustrates the measured results of a single test considering 106~PRBs. The average rate over the 10 transmissions is 135.69~Mbps. Considering a 99\% confidence level, the lower and upper limits are 135.57 and 135.79~Mbps, respectively. The peak transmission rate reached is 147.50~Mbps. Those numbers are evidence of the experiment's stability.

\begin{figure}[h!]
    \centering
    \includegraphics[width=1\linewidth]{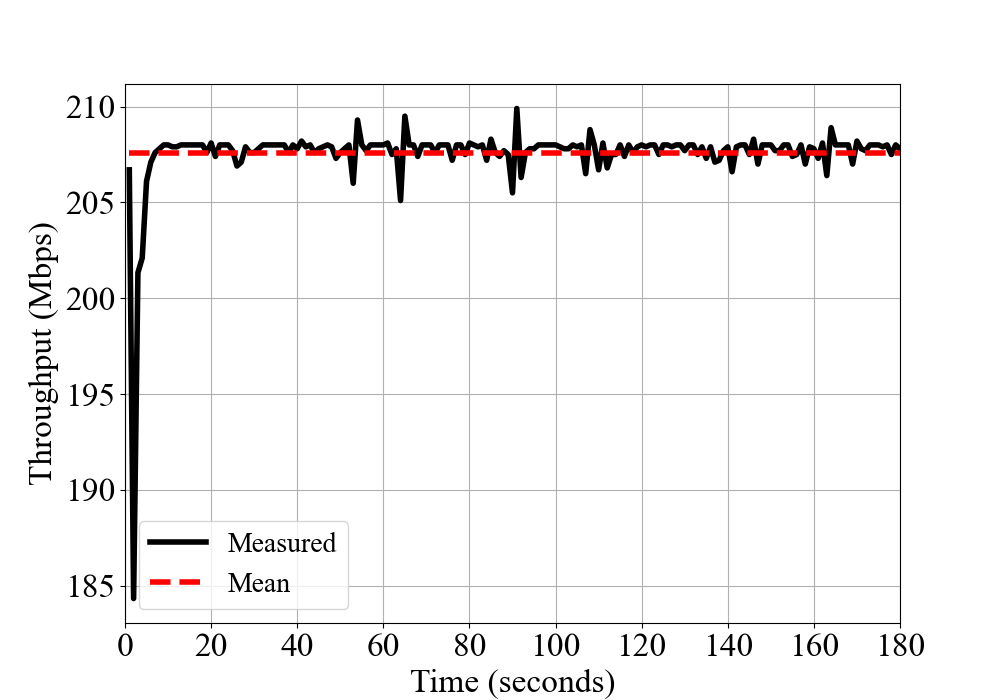}
    \caption{Throughput measurement and mean for 162 PRBs.}
    \label{fig:fig05}
\end{figure}

Figure \ref{fig:fig05} shows the measurement results considering 162~PRBs. The average rate is 207.66~Mbps. Considering a 99\% confidence level, the lower and upper limits are 207.52 and 207.66~Mbps, respectively. The peak transmission rate reached is 215.40~Mbps.

\begin{figure}[h!]
    \centering
    \includegraphics[width=1\linewidth]{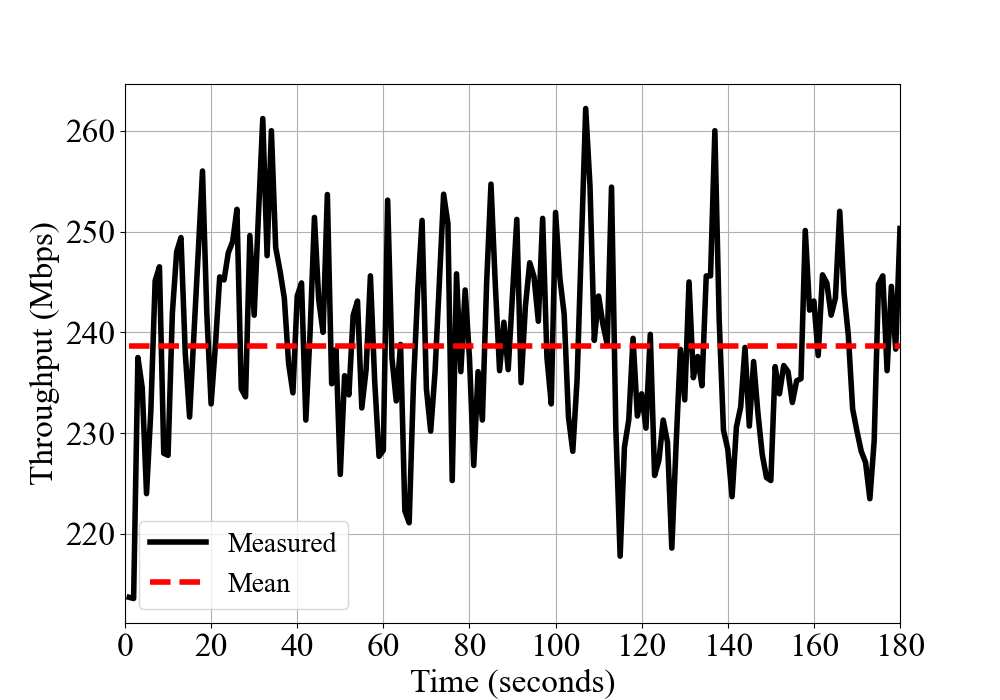}
    \caption{Throughput measurement and mean for 273 PRBs.}
    \label{fig:fig06}
\end{figure}

Figure \ref{fig:fig06} shows the results for 273 PRBs. The average rate is 238.60~Mbps, and the lower and upper limits are 227.86 and 249.33~Mbps, respectively. The peak transmission rate reached is 297.80~Mbps.

Table \ref{tab:tab01} summarizes the rates of all experiments carried out for each PRB value regarding all 10 experiment runs and considering z-core criterion to remove outliers. For 106 PRBs, the maximum rate achieved is 147.5~Mbps, while the average is 135.69~Mbps, with a standard deviation of 1.78~Mbps. As mentioned before, this shows the stability of the transmission rate during the tests. For 162 PRBs, the maximum rate achieved is 215.40~Mbps, while the average is 207.66~Mbps, with a standard deviation of 1.98~Mbps, showing similar stability in the transmission rate if compared to 106 PRBs. For 273 PRBs, the maximum rate achieved is 297.80~Mbps, with an average of 238.60~Mbps and a standard deviation of 29.02~Mbps, showing the lowest stability out of the three values of PRBs. As expected, more bandwidth consumes more computational resources, and also brings up the USRP capacity at the maximum. For 100~MHz configuration, the USRP works next to its sampling rate capacity.

\begin{table}[h!]
    \centering
    \caption{Throughput measurements for the three values of PRBs.}
    \begin{tabular}{p{1cm}p{2cm}p{2cm}p{2cm}}
    \hline
        PRB & Max (Mbps) & Mean (Mbps) & Std (Mbps)\\\hline
        106 & 147.50 & 135.69 & 1.78 \\
        162 & 215.40 & 207.66 & 1.98 \\
        273 & 297.80 & 238.60 & 29.02 \\\hline       
    \end{tabular}
    \label{tab:tab01}
\end{table}

\begin{figure}[h!]
    \centering
    \includegraphics[width=1\linewidth]{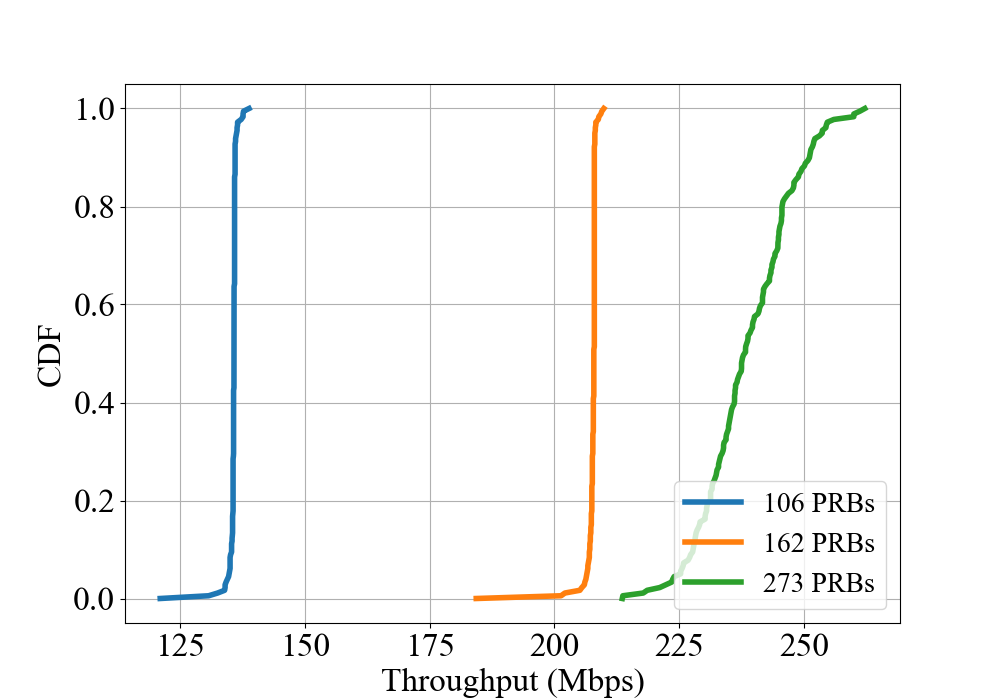}
    \caption{Cumulative distribution function for the Throughput.}
    \label{fig:fig07}
\end{figure}

Figure \ref{fig:fig07} shows the Cumulative Distribution Function (CDF) of the throughput for the three bandwidths tested. The CDFs are plotted with 1800 samples from the 10 experiments of each PRB configuration. The plot lines confirm the lower throughput variation for the narrower bandwidth, with the curve more upright for 106 and 162 PRBs.

Additionally, we measured the RTT at four different bandwidth values using the traditional ping command. The results of this experiment are summarized in Table~\ref{tab_ping_1_UE_ideal}, showing the maximum, mean, and standard deviation of RTT for each bandwidth.

The maximum RTT values increased with the number of PRBs. This could be due to higher data throughput requiring more processing time and resources, leading to increased delays. The mean RTT values also show a slight increasing trend with increasing number of PRBs. Although the increase is not as pronounced as with the maximum RTT, it suggests a consistent rise in average latency as bandwidth allocation grows. This could reflect the processing requirements scaling with bandwidth.

\begin{table}[h]
    \centering
    \caption{Ping tests for the three values of PRBs.}
    \label{tab_ping_1_UE_ideal}
    \begin{tabular}{p{1cm}p{2cm}p{2cm}p{2cm}}
    \hline
    PRB & Max (ms) & Mean (ms) & Std (ms) \\
    \hline
    106 & 28.00 & 11.98 & 3.71 \\
    162 & 35.00 & 12.35 & 4.54 \\
    273 & 79.20 & 14.03 & 10.42 \\
    \hline
    \end{tabular}
\end{table}

\begin{figure}[h!]
    \centering
    \includegraphics[width=1\linewidth, trim= 0.5cm 0cm 1cm 0cm, clip]{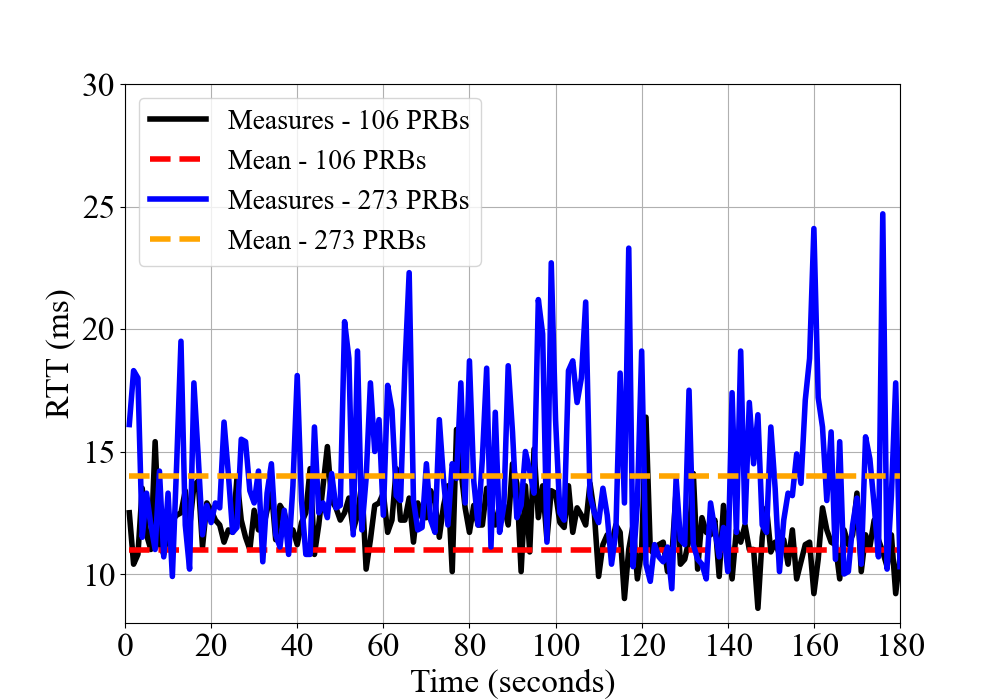}
    \caption{RTT measurement and mean for 106 and 273 PRBs.}
    \label{fig_two_PRBs}
\end{figure}

The standard deviation of RTT, which indicates the variability or consistency of the RTT measurements, shows a notable increase with higher PRBs, as illustrated in Figure~\ref{fig_two_PRBs}.

At 106 PRBs, the standard deviation is 3.71 ms and then jumps to 10.42 ms at 273 PRBs. The growing standard deviation implies that higher bandwidth allocations not only increase the mean and maximum RTT but also introduce greater variability in latency.

Finally, we report that packet loss across all experiments was zero, indicating that the throughput and RTT values were unaffected by network congestion.

\subsubsection{Two UEs Evaluation Scenario}

\begin{figure}[h!]
    \centering
    \includegraphics[width=1\linewidth]{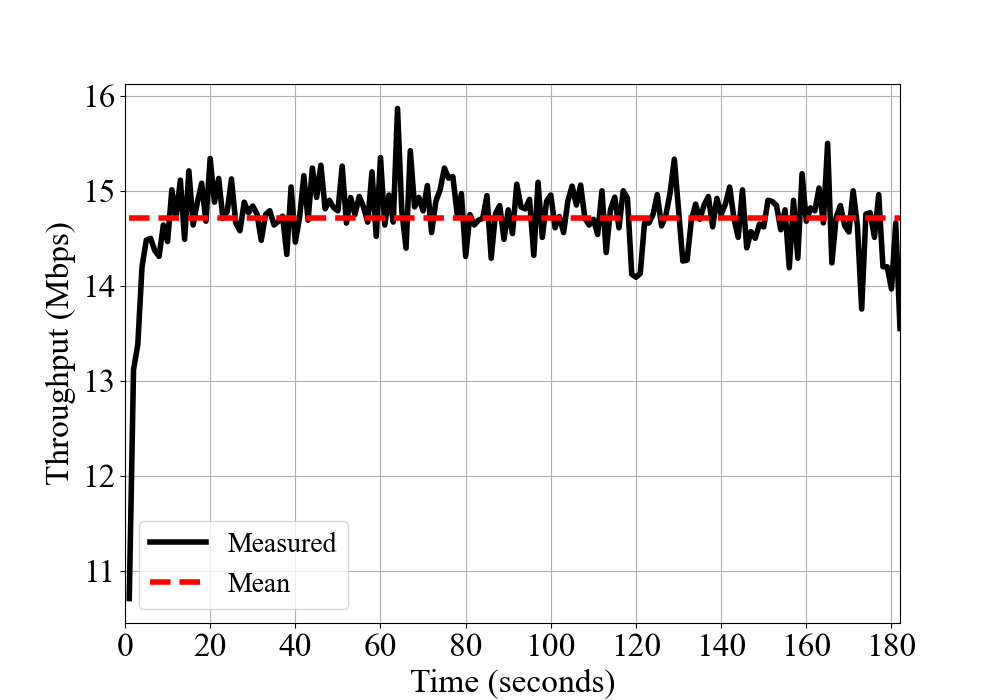}
    \caption{UE-to-UE throughput measurement and mean for 106 PRBs.}
    \label{fig_two_ue_tput}
\end{figure}

We conducted experiments to measure the throughput from two UEs connected to the same gNB. We now have two different scenarios: gNB-to-UE (with results shown in Section~\ref{sec_ideal}) and UE-to-UE (results of this section). The results from our measurements for 106 PRBs are introduced in Figure~\ref{fig_two_ue_tput} and summarized in Table~\ref{tab_2UEs}.

Considering the maximum throughput, the maximum achievable throughput in the UE-gNB scenario is significantly higher at 245~Mbps compared to 16.45~Mbps in the UE-UE scenario. This substantial difference suggests that the uplink direction limits data rates of the UE-UE scenario.

The mean throughput in the gNB-to-UE scenario is 135.69~Mbps, indicating robust data transmission capabilities between gNB and UE. Conversely, the mean throughput in the UE-to-UE scenario is 14.75 Mbps, which, while lower compared to UE-to-gNB, still demonstrates effective communication capabilities, but now depending on the ratio of UL/DL TDD slots.

\begin{table}[h!]
    \centering
    \caption{Throughput measurements for two experiment situations.}
    \label{tab_2UEs}
    \begin{tabular}{p{1cm}p{2cm}p{2cm}p{2cm}}
    \hline
        Situation & Max (Mbps) & Mean (Mbps) & Std (Mbps)\\\hline
        gNB-UE & 147.50 & 135.69 & 1.78\\
        UE-UE & 13.29 & 9.90 & 0.88 \\ \hline
    \end{tabular}
\end{table}

\begin{figure}[h!]
    \centering
    \includegraphics[width=1\linewidth, trim= 0.5cm 0cm 1cm 0cm, clip]{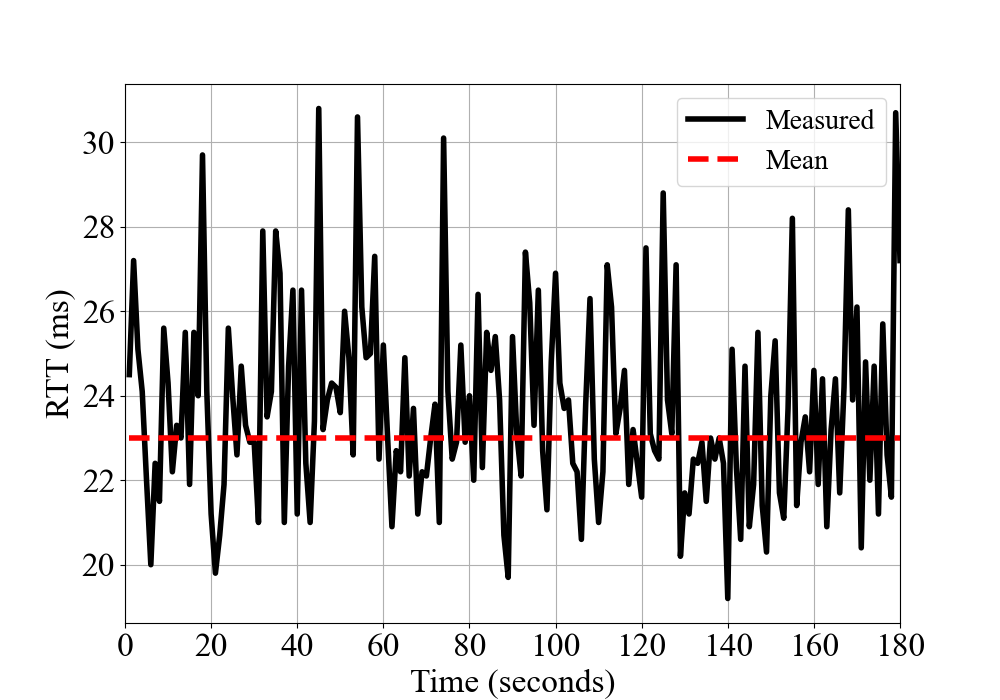}
    \caption{UE-to-UE RTT measurement and mean for 106 PRBs.}
    \label{fig_RTT_106_UE_UE}
\end{figure}

We also measure the RTT in the two above-mentioned situations. The results, highlighting the maximum, mean, and standard deviation of RTT, are summarized in Table~\ref{tab_2UEs_RTT}.

The maximum and mean RTTs are significantly higher (more than double) in UE-to-UE scenario when compared to gNB-to-UE one. This stark difference highlights that the latency can reach much higher peaks when data is transmitted between two UEs using the gNB as an intermediary.

Finally, the greater variability in RTT for UE-to-UE scenario suggests that not only is the latency higher on average, but it is also less predictable. This scenario involves bidirectional transmission between the two UEs, encompassing both uplink and downlink paths, along with additional processing at the gNB to manage data forwarding between them.

\begin{table}[h!]
    \centering
    \caption{RTT measurements for two experiment situations.}
    \label{tab_2UEs_RTT}
    \begin{tabular}{p{1cm}p{2cm}p{2cm}p{2cm}}
    \hline
        Situation & Max (ms) & Mean (ms) & Std (ms)\\\hline
        gNB-UE & 28.00 & 11.98 & 3.71\\
        UE-UE & 74.10 & 23.68 & 7.75\\ \hline
    \end{tabular}
\end{table}

\subsubsection{Distance Evaluation Scenario}

Now, we investigate the throughput performance of our 5G network at varying distances from the base station. The results are summarized in the Table~\ref{tab_tput_distance}.

As distance increases, there is a noticeable decrease in the mean throughput. At 12~m, the mean throughput further decreases to 102.35~Mbps, highlighting a more significant reduction in the average throughput over longer distances.

We observe the consistency of maximum bandwidth, even though the maximum achievable bandwidth varies across all distances, ranging between 119~Mbps to 153~Mbps. This suggests that the network can support high data rates even at the highest distance we considered.

\begin{table}[h]
    \centering
    \caption{Throughput measurements versus gNB-UE distance.}
    \label{tab_tput_distance}
    \begin{tabular}{p{2cm}p{1.5cm}p{1.5cm}p{1.5cm}}
    \hline
    \makecell{Distances \\ (meters)} & \makecell{Max BW \\ (Mbps)} & \makecell{Mean BW \\ (Mbps)} & \makecell{Std BW \\(Mbps)} \\
    \hline
    1.00 & 136.00 & 135.76 & 0.67 \\
    4.00 & 153.00 & 134.40 & 4.17 \\
    8.00 & 138.00 & 131.80 & 6.59 \\
    12.00 & 119.00 & 102.35 & 7.58 \\
    \hline
    \end{tabular}
\end{table}

\begin{figure}[h!]
    \centering
    \includegraphics[width=1\linewidth, trim= 0.5cm 0cm 1cm 0cm, clip]{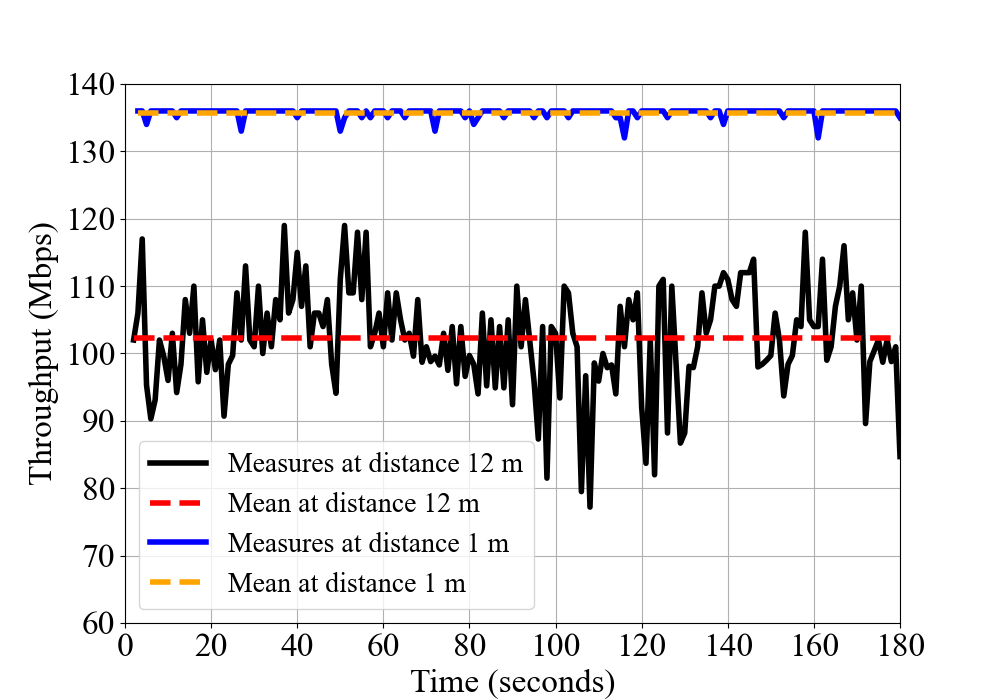}
    \caption{Throughput measurement and mean at 4~m and 12~m of gNB-UE separation.}
    \label{fig_two_distances}
\end{figure}

The standard deviation of throughput increases with distance, indicating greater variability in data rates as the distance from the base station increases, as illustrated in Figure~\ref{fig_two_distances}. We clearly see a low variation for samples at 1~m compared to samples at 12~m.
These results suggest that while the 5G network maintains high maximum bandwidth capabilities across various distances, there is a noticeable decline in mean throughput and an increase in variability (standard deviation) as distance increases.

\section{Conclusions and Future Works}
\label{sec_conclusion}

This paper presents instructions and parameter configuration of a 5G virtualized network using OpenAirInterface. We also propose three use cases, exploring a network perspective (with a spectrum-related evaluation) and a user perspective (with the throughput and ping evaluations).     

The findings of this study demonstrate the stability and potential of OAI code for prototyping 5G enhancements, including their evaluation in a real low-cost platform. Thus, industry-leading vendors, small/medium companies, start-ups, and academic institutions could explore the OAI from training to conceptual initiative.  

Measurements from a multicell environment are a key next step, broadening the analysis to include cochannel interference and other user-related metrics such as latency and error rate. 

\section*{Declarations}

\begin{acknowledgements}
The authors gratefully acknowledge Lenovo Brazil and UFRN for their technical and financial support.  
\end{acknowledgements}

\begin{funding}
This research was partially funded by Lenovo, as part of its R\&D investment under Brazilian Informatics Law, and by the \textit{Coordenacao de Aperfeicoamento de Pessoal de Nivel Superior - Brasil} (CAPES) - Finance Code 001.
\end{funding}

\begin{contributions}
AC, VS, IR, MF and AN contributed to the conception and validation of this study. MD, AC, NO, PE, PF, CL, JG, DL area responsible for methodology definition and experiments realization. MD, VS, AC, NO, PF, CL, JG, DL, IR are this manuscript’s main contributors and writers. AC, AN, MF, VS supervised all experiments and are responsible for the final version of paper. All authors read and approved the final manuscript.
\end{contributions}

\begin{interests}
The authors declare that they have no competing interests.
\end{interests}

\begin{materials}
The datasets generated and analyzed during the current study and a tutorial material to reproduce all results are available at \url{https://github.com/lance-ufrn/oai_bench_tuto}.
\end{materials}

\bibliographystyle{apalike-sol}
\bibliography{refs}

@techreport{3gpp_38.104,
 author = {3GPP},
 institution = {{3rd Generation Partnership Project (3GPP)}},
 number = {38.104},
 title = {{NR; Base Station (BS) radio transmission and reception}},
 type = {Technical Specification (TS) - Release 15},
 year = {2024},
 note = {\url{https://www.3gpp.org/ftp/Specs/archive/38_series/38.104/38104-fk0.zip}}
}

@techreport{3gpp_38.101-1,
 author = {3GPP},
 institution = {{3rd Generation Partnership Project (3GPP)}},
 number = {38.101-1},
 title = {{NR; User Equipment (UE) radio transmission and reception; Part 1: Range 1 Standalone}},
 type = {Technical Specification (TS) - Release 15},
 year = {2024},
 note = {\url{https://www.3gpp.org/ftp/Specs/archive/38_series/38.101-1/38101-1-fr0.zip}}
}

@techreport{3gpp.38.855,
 author = {3GPP},
 institution = {{3rd Generation Partnership Project (3GPP)}},
 number = {38.855},
 title = {{Study on NR positioning support}},
 type = {Technical report (TR) - Release 16},
 year = {2019},
 note = {\url{https://www.3gpp.org/ftp//Specs/archive/38_series/38.855/38855-g00.zip}}
}

@article{KALTENBERGER2020107284,
author = {Florian Kaltenberger and Aloizio P. Silva and Abhimanyu Gosain and Luhan Wang and Tien-Thinh Nguyen},
title = {{OpenAirInterface: Democratizing innovation in the 5G Era}},
journal = {Computer Networks},
volume = {176},
pages = {107284},
year = {2020},
issn = {1389-1286},
doi = {https://doi.org/10.1016/j.comnet.2020.107284},
}

@INPROCEEDINGS{9310895,
  author={Wiranata, Fandi A. and Shalannanda, Wervyan and Mulyawan, Rahmat and Adiono, Trio},
  booktitle={2020 14th International Conference on Telecommunication Systems, Services, and Applications (TSSA}, 
  title={{Automation of Virtualized 5G Infrastructure Using Mosaic 5G Operator over Kubernetes Supporting Network Slicing}}, 
  year={2020},
  pages={1-5},
  doi={10.1109/TSSA51342.2020.9310895}
}

@inproceedings{Nvoa2020RANSU,
  author={Lucas N{\'o}voa and Virg{\'i}nia Tavares and Cleverson Veloso Nahum and Pedro Batista and Aldebaro Klautau},
  title={{RAN Slicing using OpenAirInterface and FlexRAN in a Virtualized Scenario}},
  year={2020},
  url={https://api.semanticscholar.org/CorpusID:229331076}
}

@INPROCEEDINGS{9318327,
  author={Manco, Julio and Baños, Guillermo Gallud and Härri, Jérôme and Sepulcre, Miguel},
  booktitle={2020 IEEE Vehicular Networking Conference (VNC)}, 
  title={{Prototyping V2X Applications in Large-Scale Scenarios using OpenAirInterface}}, 
  year={2020},
  pages={1-4},
  doi={10.1109/VNC51378.2020.9318327}
}

@INPROCEEDINGS{9154290,
  author={Magounaki, Theoni and Kaltenberger, Florian and Knopp, Raymond},
  booktitle={2020 IEEE 21st International Workshop on Signal Processing Advances in Wireless Communications (SPAWC)}, 
  title={{Modeling the Distributed MU-MIMO OAI 5G testbed and group-based OTA calibration performance evaluation}}, 
  year={2020},
  pages={1-5},
  doi={10.1109/SPAWC48557.2020.9154290}
}

@INPROCEEDINGS{9633433,
  author={Gao, Yuehong and Zhang, Xiaohang and Yuan, Hongquan},
  booktitle={2021 IEEE 3rd International Conference on Civil Aviation Safety and Information Technology (ICCASIT)}, 
  title={{Integration and Connection Test for OpenAirInterface 5G Standalone System}}, 
  year={2021},
  pages={1010-1014},
  doi={10.1109/ICCASIT53235.2021.9633433}
}

@INPROCEEDINGS{9739187,
  author={Nakkina, Sai Shruthi and Balijepalli, Sudhakar and Murthy, Chandra R.},
  booktitle={WSA 2021; 25th International ITG Workshop on Smart Antennas}, 
  title={{Performance Benchmarking of the 5G NR PHY on the OAI Codebase and USRP Hardware}}, 
  year={2021},
  pages={1-6},
  doi={}
}

@INPROCEEDINGS{9442028,
  author={Vilakazi, Mla and Burger, Chris R. and Mboweni, Lawrence and Mamushiane, Lusani and Lysko, Albert A.},
  booktitle={2021 7th International Conference on Advanced Computing and Communication Systems (ICACCS)}, 
  title={{Evaluating an Evolving OAI Testbed: Overview of Options, Building tips, and Current Performance}}, 
  year={2021},
  volume={1},
  pages={818-827},
  doi={10.1109/ICACCS51430.2021.9442028}
}

@ARTICLE{9610039,
  author={Papatheofanous, Elissaios Alexios and Reisis, Dionysios and Nikitopoulos, Konstantinos},
  journal={IEEE Access}, 
  title={{LDPC Hardware Acceleration in 5G Open Radio Access Network Platforms}}, 
  year={2021},
  volume={9},
  pages={152960-152971},
  doi={10.1109/ACCESS.2021.3127039}}

@INPROCEEDINGS{9771860,
  author={Atalay, Tolga O. and Stojadinovic, Dragoslav and Stavrou, Angelos and Wang, Haining},
  booktitle={2022 IEEE Wireless Communications and Networking Conference (WCNC)}, 
  title={{Scaling Network Slices with a 5G Testbed: A Resource Consumption Study}}, 
  year={2022},
  pages={2649-2654},
  doi={10.1109/WCNC51071.2022.9771860}}

@INPROCEEDINGS{9880817,
  author={Li, Dexin and Chu, Xinghe and Wang, Luhan and Lu, Zhaoming and Zhou, Shuya and Wen, Xiangming},
  booktitle={2022 IEEE/CIC International Conference on Communications in China (ICCC)}, 
  title={{Performance Evaluation of E-CID based Positioning on OAI 5G-NR Testbed}}, 
  year={2022},
  pages={832-837},
  doi={10.1109/ICCC55456.2022.9880817}
}

@INPROCEEDINGS{9771597,
  author={Ursu, Răzvan-Mihai and Papa, Arled and Kellerer, Wolfgang},
  booktitle={2022 IEEE Wireless Communications and Networking Conference (WCNC)}, 
  title={{Experimental Evaluation of Downlink Scheduling Algorithms using OpenAirInterface}}, 
  year={2022},
  pages={84-89},
  doi={10.1109/WCNC51071.2022.9771597}
}

@INPROCEEDINGS{10028998,
  author={Dludla, Gcina and Vilakazi, Mla and Burger, Chris R. and Lysko, Albert A. and Ngcama, Lwando and Mboweni, Lawrence and Masonta, Moshe and Kobo, Hlabi and Mamushiane, Lusani},
  booktitle={2022 5th International Conference on Multimedia, Signal Processing and Communication Technologies (IMPACT)}, 
  title={{Testing Performance for Several Use Cases with an Indoor OpenAirInterface USRP-Based Base Station}}, 
  year={2022},
  pages={1-5},
  doi={10.1109/IMPACT55510.2022.10028998}
}

@INPROCEEDINGS{10170129,
  author={D., Bhavana and Chaudhari, Shilpa},
  booktitle={2023 4th International Conference for Emerging Technology (INCET)}, 
  title={{Performance Evaluation of Openairinterface's Scheduling Algorithms for 5G Networks}}, 
  year={2023},
  pages={1-4},
  doi={10.1109/INCET57972.2023.10170129}
}

@INPROCEEDINGS{10118776,
  author={Sahbafard, Arash and Schmidt, Robert and Kaltenberger, Florian and Springer, Andreas and Bernhard, Hans-Peter},
  booktitle={2023 IEEE Wireless Communications and Networking Conference (WCNC)}, 
  title={{On the Performance of an Indoor Open-Source 5G Standalone Deployment}}, 
  year={2023},
  pages={1-6},
  doi={10.1109/WCNC55385.2023.10118776}
}

@INPROCEEDINGS{10041332,
  author={Thakkar, Priyal and Sanadhya, Shashvat and Gandotra, Pimmy and Lall, Brejesh},
  booktitle={2023 15th International Conference on COMmunication Systems \& NETworkS (COMSNETS)}, 
  title={{A 5G OpenAirInterface (OAI) Testbed with MEC: Deployment, Application testing and Slicing Support}}, 
  year={2023},
  pages={757-762},
  doi={10.1109/COMSNETS56262.2023.10041332}
}

@misc{devicespecifications-moto-g50-specification,
  title = {{Motorola Moto G50 5G - Specifications}},
  author = {{Device Specifications}},
  year = {2021},
  url = {https://www.devicespecifications.com/en/model/2fc15727},
  note = {Access: dez/2023}
}

@misc{phonemore-moto-g50-specification,
  title = {{Motorola Moto G50 5G: All Specs}},
  author = {{PHONEmore}},
  year = {2021},
  url = {https://www.phonemore.com/specs/motorola/moto-g50-5g/},
  note = {Access: dez/2023}
}

@misc{oai-oam,
title = {{OAI Operations and Maintenance (OAM) project group }},
author = {{OSA}},
year = {2024},
note = {https://openairinterface.org/projects/oam-project-group/, access in feb/2024}
}

@misc{oai-cn,
author = {{OSA}},
title = {{OAI-5G Core Network (CN) project group}},
year = {2024},
note = {https://openairinterface.org/oai-5g-core-network-project/, access in feb/2024}
}

@misc{oai-ran,
author = {{OSA}},
title = {{OAI 5G RAN Project Group}},
year = {2024},
note = {https://openairinterface.org/oai-5g-ran-project/, access: feb/2024}
}

@misc{oai-about,
title = {{About the OpenAirInterface Software Alliance}},
author = {{OSA}},
year = {2024},
note = {https://openairinterface.org/about-us/, access: feb/2024}
}

@misc{upf-vpp-oai-git,
title = {{An implementation of the 5G Core network by the OpenAirInterface community}},
author = {{OSA}},
year = {2024},
note = {https://gitlab.eurecom.fr/oai/cn5g/oai-cn5g-upf-vpp, access: mar/2024}
}

@misc{upf-vpp-git,
title = {{Vector Packet Processing Platform}},
author = {{FDio}},
year = {2024},
note = {https://github.com/fdio/vpp, access: mar/2024}
}

@misc{bubble-max-gain-rx,
title = {{Specific Parameters per RU/split}},
author = {{Mohammadi}},
year = {2024},
note = {https://bubbleran.com/docs/devops/config/oai-ran/, access: mar/2024}
}

\end{document}